\documentclass[aps,twocolumn,showpacs,superscriptaddress,floatfix,nofootinbib]{revtex4-2}

\pdfoutput=1
\usepackage[utf8]{inputenc}
\usepackage[english]{babel}
\usepackage[T1]{fontenc}
\usepackage{csquotes}
\usepackage{physics}

\usepackage{amsmath,amsfonts}

\usepackage[dvipsnames]{xcolor}

\usepackage[colorlinks=true]{hyperref}
\usepackage[capitalise]{cleveref}

\usepackage{listings}
\usepackage{parskip}

\usepackage{graphicx}
\graphicspath{{./}}

\usepackage{qcircuit}

\usepackage{bm}

\begin{document}
\title{Adaptive projected variational quantum dynamics}

\author{David Linteau}
\email{david.linteau@epfl.ch}
\affiliation{Institute of Physics, \'{E}cole Polytechnique F\'{e}d\'{e}rale de Lausanne (EPFL), CH-1015 Lausanne, Switzerland}
\affiliation{Center for Quantum Science and Engineering, \'{E}cole Polytechnique F\'{e}d\'{e}rale de Lausanne (EPFL), CH-1015 Lausanne, Switzerland}

\author{Stefano Barison}
\affiliation{Institute of Physics, \'{E}cole Polytechnique F\'{e}d\'{e}rale de Lausanne (EPFL), CH-1015 Lausanne, Switzerland}
\affiliation{Center for Quantum Science and Engineering, \'{E}cole Polytechnique F\'{e}d\'{e}rale de Lausanne (EPFL), CH-1015 Lausanne, Switzerland}
\affiliation{National Centre for Computational Design and Discovery of Novel Materials MARVEL, EPFL, Lausanne, Switzerland}

\author{Netanel H. Lindner}
\affiliation{Physics Department, Technion, 320003 Haifa, Israel}

\author{Giuseppe Carleo}
\affiliation{Institute of Physics, \'{E}cole Polytechnique F\'{e}d\'{e}rale de Lausanne (EPFL), CH-1015 Lausanne, Switzerland}
\affiliation{Center for Quantum Science and Engineering, \'{E}cole Polytechnique F\'{e}d\'{e}rale de Lausanne (EPFL), CH-1015 Lausanne, Switzerland}
\affiliation{National Centre for Computational Design and Discovery of Novel Materials MARVEL, EPFL, Lausanne, Switzerland}

\begin{abstract}
We propose an adaptive quantum algorithm to prepare accurate variational time evolved wave functions.
The method is based on the projected Variational Quantum Dynamics (pVQD) algorithm, that performs a global optimization with linear scaling in the number of variational parameters.
Instead of fixing a variational ansatz at the beginning of the simulation, the circuit is grown systematically during the time evolution.
Moreover, the adaptive step does not require auxiliary qubits and the gate search can be performed in parallel on different quantum devices.
We apply the new algorithm, named Adaptive pVQD, to the simulation of driven spin models and fermionic systems, where it shows an advantage when compared to both Trotterized circuits and non-adaptive variational methods.
Finally, we use the shallower circuits prepared using the Adaptive pVQD algorithm to obtain more accurate measurements of physical properties of quantum systems on hardware. 
\end{abstract}

\maketitle

\section{Introduction}
\label{sec:intro}

Simulation of static and dynamic properties of quantum systems is a notoriously hard task for classical computers.
While analytical solutions are available only for specific cases, the amount of time and computing resources required in general by exact numerical methods is exponential in the system size, making the calculations quickly unfeasible. 
While several approximated many-body numerical techniques have been proposed \cite{2010_sandvik_computational_methods, 2017_carleo_nqs, 2019_orus_tensor_networks_review, 2019_carleo_ml_in_physics}, the accurate description of important physical and chemical phenomena is a very active research problem \cite{Cady2008,Schimka2010,Leggett2006,Balents2010}.

In recent years, quantum computers have seen significant developments \cite{2019_arute_quantum_supremacy, 2020_zhong_quantum_advantage_photons, 2022_huang_quantum_advantage}, opening potential opportunities for scientific discoveries.
Hardware capabilities continue to advance steadily, and we can already create and manipulate complex many-body quantum systems \cite{2014_dolde_optimal_control, 2014_waldherr_optimal_control, 2019_nam_ionq_progress, 2020_wan_boulder_ion_traps, 2020_hughes_high_fidelity_ent_gate,Kim2023}.
However, large-scale fault-tolerant quantum computers remain far in the future, and contemporary devices show limitations in connectivity, size, and coherence times.

Accounting for these constraints, Variational Quantum Algorithms (VQAs) have emerged as the leading strategy to take advantage of near-term quantum devices \cite{2021_cerezo_vqa_rev, 2022_bharti_nisq_algo_rev,2019_yuan_variational_simulations_rev,2014_peruzzo_vqe}.
In this class of algorithms, the solution of a given problem (e.g. finding the ground state of a physical system) is encoded in a quantum circuit that depends on some parameters optimized with the aid of a classical device.
VQAs have not only been proposed for quantum simulations but also for a variety of different applications, such as machine learning \cite{2017_biamonte_quantum_machine_learning, 2019_cong_quantum_cnn}, combinatorial optimization \cite{2014_farhi_quantum_combinatorial_opt, 2018_wang_qaoa_maxcut}, quantum error correction \cite{2017_johnson_var_error_correction, 2021_xu_var_error_correction} and compilation \cite{2019_khatri_var_quantum_compiling, 2020_sharma_var_quantum_compiling, 2022_jones_var_quantum_compilation}.
Variational schemes have also been introduced in quantum dynamics \cite{2017_ying_vqa_example, 2020_crstoiu_vff, 2021_yao_adaptive_tdva,Hsuan_2021, Barratt_2021, 2021_barison_p-vqd, 2022_berthusen_dynamics_on_hardware, 2022_barison_vfk, Miessen2023_te}, as a more efficient alternative to Trotterization \cite{1959_trotter, 1976_suzuki, 1997_abrams_trotter_for_simulation, 2001_ortiz_trotter_for_simulation,Tacchino_2019}. 
The accuracy of a variational quantum simulation is then tied to the ability of a parameterized circuit to describe time-evolved wave functions.
Even if the initial wave function is well-described by the chosen parameterized circuit, the complexity of the time-evolved wave functions varies with time  and the chosen circuit may fail to describe them.
The choice of the parameterized circuit is therefore crucial and it remains an open problem in variational simulations of quantum dynamics.

Adaptive schemes have been proposed in the context of variational ground state search \cite{2019_grimsley_adapt-vqe, 2021_tang_qubit_adapt_vqe, 2022_van_dyke_pool_tiling, 2022_economou_tetris} especially to avoid committing to a particular parameterized circuit.
The key idea is to construct the parameterized circuit during optimization.
By systematically appending specific quantum gates to the parameterized circuit, adaptive methods have been shown to surpass standard approaches in the number of operations required and in the accuracy of the final results.
Moreover, adaptive methods provide flexible circuits suited for dynamics simulations \cite{2021_yao_adaptive_tdva, Niladri_2023}.
However, including an adaptive step for dynamics usually requires measurements of additional quantities, that might be difficult to perform, or auxiliary qubits.

In this work, we introduce an adaptive variational algorithm for real-time evolution based on the projected Variational Quantum Dynamics (pVQD) algorithm \cite{2021_barison_p-vqd}, denoted Adaptive pVQD.
The method inherits all the properties of the original pVQD algorithm and integrates the adaptive modification of the parameterized circuit without requiring auxiliary qubits. 
The structure of this paper is as follows: in \cref{sec:method} we present the algorithm and describe how the adaptive routine is performed; in \cref{sec:results} we apply the method to study a time-dependent and a fermionic system, benchmarking the method against Trotter evolution and the original pVQD algorithm;
\cref{sec:conclusions} concludes the paper with some considerations and outlooks on the proposed method.

\section{Method}
\label{sec:method}

\begin{figure*}
    \centering
    \includegraphics[width=\textwidth]{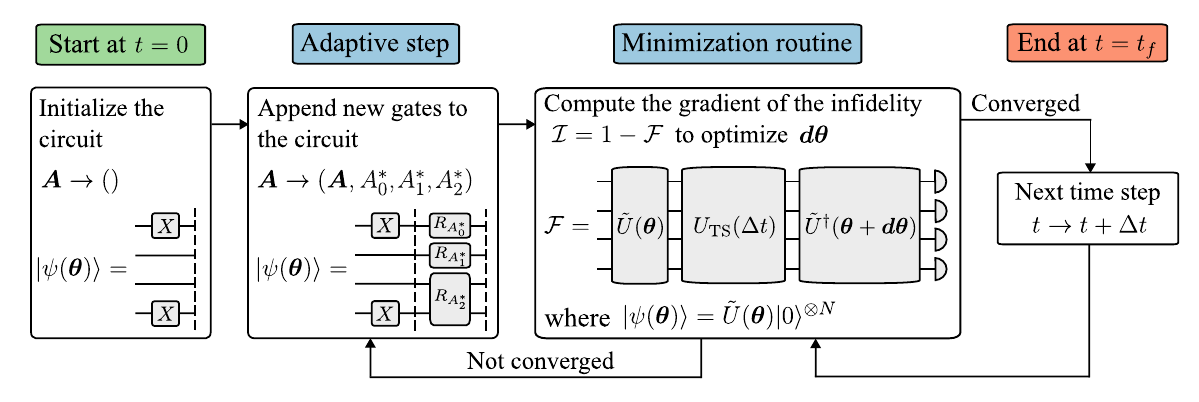}
    \caption{Flowchart of the time evolution of the Adaptive pVQD algorithm.
    Starting with a parameter-free circuit, we discretize the time evolution into multiple time steps.
    At each time step we optimize the parameters to approximate the real time evolution of the quantum system.
    If the optimization does not converge to the required accuracy, or the ansatz does not contain any parameter, then rotations $\{R_{A_i^*}\}$ based on the generators $\{A_i^*\}$ are appended to the circuit according to the adaptive step procedure described in \cref{subsec:Adaptive step}. 
    The algorithm stops once the final time $t_f$ is reached.
    }
    \label{fig:flowchart_algorithm}
\end{figure*}

Consider a physical system governed by a Hamiltonian $H$.
For clarity of exposition, we focus on time-independent Hamiltonians.
However, this is not a requirement of the algorithm, as we explicitly show in \cref{sec:results}.
To simulate the dynamics of quantum systems on a quantum computer, we have to prepare the time-evolved wave function $|\Psi(t) \rangle = U(t)|\psi_0 \rangle $, where $| \psi_0 \rangle = U_0 |0 \rangle^{\otimes N}$ is the initial state, $N$ indicates the number of qubits representing the physical system and $U(t)$ is the unitary time evolution operator.
The Adaptive pVQD algorithm aims to approximate the state $|\Psi(t) \rangle$ using parameterized states of the form

\begin{equation} \label{eq:parameterized_state}
    | \psi (\bm{\theta}, \bm{A}) \rangle = U(\bm{\theta}, \bm{A}) | \psi_0 \rangle = \prod_{i} e^{-i \theta_i A_i} | \psi_0 \rangle,
\end{equation}
where each real parameter $\theta_i \in \bm{\theta}$ is associated to a Hermitian generator $A_i \in \bm{A}$.
The parameterized state is therefore specified by the set of parameters and operators $\{ \bm{\theta}, \bm{A} \}$, and it can be implemented as a quantum circuit. 
From now on, we adopt the notation $| \psi(\bm{\theta}) \rangle \equiv | \psi(\bm{\theta}, \bm{A}) \rangle $ and $U(\bm{\theta}) \equiv U(\bm{\theta}, \bm{A})$. 

To simulate a physical model until a final time $t_f$, we divide the evolution into small time intervals $\Delta t$. 
We further assume that the parameterized state $| \psi  ( \bm{\theta} ) \rangle$ is a good approximation of the time-evolved wave function at time $t$.
The wave function at time $t + \Delta t$ can thus be represented by $ U_\text{TS}(\Delta t) | \psi  ( \bm{\theta} ) \rangle $, where $U_\text{TS}(\Delta t)$ is a Trotter-Suzuki decomposition of the time evolution operator $U(\Delta t)$ \cite{1959_trotter,1976_suzuki}.
In this manuscript we use a first order decomposition, but higher orders can be considered.
The choice of the optimal $\Delta t$ is problem dependent and will be discussed in \cref{sec:results}.
We then approximate the evolution step $t \to t+\Delta t$ using a new set of parameters $\bm{\theta} \to \bm{\theta} + \bm{d \theta}$ that maximizes the overlap between  $ U_\text{TS}(\Delta t) | \psi  ( \bm{\theta} ) \rangle $ and $| \psi( \bm{\theta}+\bm{d \theta}) \rangle$.
This can be achieved by minimizing, with respect to $\bm{d \theta}$, the infidelity
\begin{equation} \label{eq:infidelity}
    \mathcal{I}(\bm{d \theta},\Delta t) = 1 - \mathcal{F}(\bm{d \theta},\Delta t),
\end{equation}
where the fidelity
\begin{equation} \label{eq:fidelity}
    \mathcal{F}(\bm{d \theta},\Delta t) = | \langle \psi( \bm{\theta}+\bm{d \theta}) | U_\text{TS} (\Delta t) | \psi  ( \bm{\theta} ) \rangle |^2
\end{equation}
can be measured on a quantum device \cite{2021_barison_p-vqd}.

At each time step, the initial parameters and operators $\{ \bm{\theta}, \bm{A} \}$ are those obtained at the previous time step.
Assuming that the set of operators $\bm{A}$ is sufficient to describe the state at time $t + \Delta t$, we find the parameter shift $\bm{d \theta}^*$ that minimizes $\mathcal{I}(\bm{d \theta},\Delta t)$.
Details about the minimization routine can be found in \cref{appendix:Minimization routine}.
If the minimization routine is not successful, new gates built using generators $(A_0^*, A_1^*, \cdots, A_k^*)$ from the operator pool are added to the parameterized circuit following the adaptive procedure described in \cref{subsec:Adaptive step}.
This adaptive procedure is repeated up until the convergence criteria are met.

The algorithm starts with the initial state $|\psi_0\rangle$ represented by an empty set of operators. As needed, new gates are added through the time evolution until the chosen final time $t_f$. 
The complete procedure is illustrated in \cref{fig:flowchart_algorithm}.
We note that the original pVQD scheme \cite{2021_barison_p-vqd} can be recovered by fixing the set of operators $\bm{A}$ through the entire simulation.

\subsection{Adaptive step} 
\label{subsec:Adaptive step}

When the parameterized circuit $| \psi(\bm{\theta}) \rangle$ is not expressive enough to accurately describe the time step evolution by only shifting the variational parameters, we add new gates to it. 
This is referred to as the adaptive step of the algorithm.
Given an operator pool, we determine the best gate to grow the quantum circuit.
As first proposed in \cite{2019_grimsley_adapt-vqe}, we look for the operator whose gate maximizes the derivative of the cost function with respect to its parameter.
This is achieved by iterating over all the operators in the pool, a step that can be performed in parallel even on different quantum devices.

For ground state methods, the cost function is the energy of the system, while the gradient is obtained by measuring the expectation value of the commutator between the trial operator and the Hamiltonian \cite{2019_grimsley_adapt-vqe,anastasiou2023really}.
We must ensure that is possible to apply a similar procedure when dynamics is considered.
In the adaptive scheme proposed in \cite{2021_yao_adaptive_tdva}, this step requires an additional measurement of the variance of the Hamiltonian with respect to the non-adaptive case.
In our method, the gradient of the fidelity with respect to the shift $d\theta_a$ of parameter $\theta_a$ associated with a trial operator $A_a$  has the form

\begin{equation} \label{eq:grad_fidelity}
    \frac{ \partial \mathcal{F} }{ \partial d\theta_a } = \bra{\phi(\bm{\theta},\Delta t )}  e^{-id\theta_a A_a} [P_0,iA_a] e^{id\theta_a A_a} \ket{\phi(\bm{\theta},\Delta t )},
\end{equation}

where we define the projector $P_0 = \dyad{\psi_0}{\psi_0}$ and the state $\ket{\phi(\bm{\theta},\Delta t )} = U^{\dagger}(\bm{\theta}) U_\text{TS}(\Delta t) \ket{\psi(\bm{\theta} )}$ (see \cref{appendix:gradient_fidelity} for the full derivation).
To ensure continuity of  time evolution, we initially set  $\theta_a=0$.
We note that measuring the derivative of the fidelity corresponds to measuring the Hermitian operator $[P_0,iA_a]$ with respect to the pVQD circuit $U^{\dagger}(\bm{\theta})U_\text{TS}(\Delta t) \ket{\psi(\bm{\theta} )}$ modified by the addition of the gate $e^{id\theta_a A_a}$.
However, we evaluate the derivative  using the parameter shift rule \cite{2021_mari_param_shift_rule}, as for the minimization routine (for more details, see \cref{appendix:Minimization routine}).
This operator search is still parallelizable on multiple devices and does not require auxiliary qubits.

The adaptive step has been lately extended and optimized \cite{2021_tang_qubit_adapt_vqe,2022_economou_tetris,anastasiou2023really}, with new protocols that greatly reduce the computational resources required with respect to the first proposal.
In particular, we adopt the scheme presented in \cite{2022_economou_tetris}, which increases the depth of the parameterized circuit $|\psi(\bm{\theta})\rangle$ by 1 at every adaptive step.
While the infidelity defined in \cref{eq:infidelity} remains above a fixed threshold $\varepsilon$, additional adaptive steps are performed.
For a detailed description, see \cref{appendix:tetris_adaptive_steps}.

\subsection{Operator pool} 
\label{sec:Operator pool}

The choice of the operator pool is a key ingredient in the success and efficiency of adaptive variational algorithms.
Having a complete pool of operators is exponentially complex in the size of the physical system, therefore, one has to make some restrictions in its selection.
Many different strategies have been proposed, such as the creation of a minimally complete pool \cite{2021_tang_qubit_adapt_vqe,shkolnikov2021avoiding}, the inclusion of symmetries directly in the operator pool \cite{2021_yordanov_adapt}, or the extension of a complete pool acting on a subsystem of the studied model \cite{2022_van_dyke_pool_tiling}.

In the study of the dynamics, we can refer to the Trotterization of the time evolution operator to select the pool.
In particular, we consider local (L) and non-local (NL) operator pools, respectively, given by
\begin{align}
    \mathcal{A}_\text{L} &= \{ X_i, Y_i, Z_i \}_{i=0}^{N-1} \label{eq:local_pool} \\
    & \ \ \ \ \cup \{ X_i X_{i+1}, Y_i Y_{i+1}, Z_i Z_{i+1} \}_{0 \le i \le N-2}, \nonumber \\
    \mathcal{A}_\text{NL} &= \{ X_i,  Y_i,  Z_i, X_i X_j, Y_i Y_j,  Z_i Z_j \}_{0 \le i < j \le N-1}, \label{eq:non_local_pool}
\end{align}
where $X_i, Y_i$ and $Z_i$ are the Pauli gates acting on site $i$.
Given that $\mathcal{A}_\text{L} \subset \mathcal{A}_\text{NL}$, we expect that $\mathcal{A}_\text{NL}$ will generate more flexible parameterized states.
However, not only the choice of $\mathcal{A}_\text{NL}$ leads to a measurement overhead, but the non-local nature of this pool may add long-range controlled-NOT (CNOT) gates to the circuit, according to the device connectivity.
In \cref{sec:results}, we report the comparison of the two pools in the study of a fermionic system.

\section{Results}
\label{sec:results}

We apply the Adaptive pVQD method to the study of the 1D Heisenberg XYZ model with an external driving field and the 2D Fermi-Hubbard model.
Both have non-trivial dynamics and open the pVQD method to the study of time-dependent and fermionic systems.
In both cases, open boundary conditions were imposed.

\subsection{Driven Heisenberg model} 
\label{sec:Driven Heisenberg model}

\begin{figure}[h]
    \centering
    \includegraphics[width=\columnwidth]{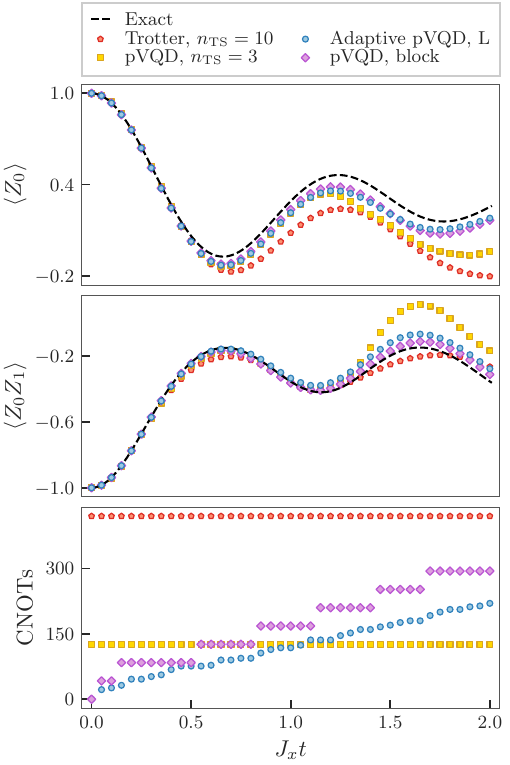}
    \caption{
    Dynamics of the driven Heisenberg XYZ model studied with the Adaptive pVQD algorithm with local pool (L), compared to standard Trotter evolution, pVQD and pVQD with block extensions.
    The plot shows the results for an open chain of $L=8$ spins with $J_x=1, J_y=0.8$ and $J_z=0.6$.
    The top and middle panel show the measurements of a single spin observable and a correlator, respectively.
    The bottom panel shows the number of CNOTs in the circuit describing the time-evolved wave function.
    The simulation started in the antiferromagnetic state $| \psi_0 \rangle = |01010101 \rangle$, and the infidelity threshold was set to $\varepsilon = 10^{-4}$ for all the variational methods.}
    \label{fig:xyz_floquet_statev}
\end{figure}

Given an open chain of $L$ spins, the driven Heisenberg XYZ Hamiltonian can be written as:

\begin{align}
 \label{eq:driven_heisenberg_model}
    H(t) =  \sum_{i=0}^{L-2} ( J_x X_i X_{i+1} + J_y Y_i Y_{i+1} + J_z Z_i Z_{i+1} ) + D(t)
\end{align}
where $J_x, J_y$ and $J_z$ are coupling parameters and $D(t)$ is the time-dependent driving term.
Many different driving terms can be applied to the system. Among those we choose 

\begin{align}
 \label{eq:driving_term}
    D(t) &=  \sum_{i=0}^{L-1} (-1)^i \sin( \omega t) Z_i \, ,
\end{align}
where $\omega$ is the driving frequency.

First, we investigate the performance of the Adaptive pVQD algorithm with a local pool on a perfect simulator and compare to Trotterized circuits and the original implementation of pVQD.
We consider $J_x=1, J_y=0.8, J_z=0.6$, an antiferromagnetic initial state $|\psi_0\rangle = |0101\rangle $ and a final evolution time $t_f = 2$.
In the classic version of the pVQD algorithm, we have to choose an ansatz for the time evolved wave function.
We consider a circuit equivalent to a Trotter step where all the rotations are defined by variational parameters.
The Trotter step circuit implementation for this model is shown in \cref{appendix:Trotter step circuits}.
Both the Trotter and the pVQD full circuits are then obtained repeating this structure $n_\text{TS}$ times.
In particular, we fix $n_\text{TS} = 10$ for the Trotter circuit and $n_\text{TS} = 3$ for the pVQD ansatz.

After running the algorithms, we compare the different circuits obtained and use them to measure expectation values of single- and two-spin observables.
The results are shown in \cref{fig:xyz_floquet_statev}.
The Trotter circuit lags behind variational methods both in terms of accuracy and resource required.
The pVQD method instead achieves accurate results until $t=1.0$, where the associated circuit becomes shallower than the one of Adaptive pVQD.
This phenomenon suggests that in that time step the fixed representation power is the main source of error in the variational calculations.

In order to show the flexibility of the Adaptive pVQD, we implement a naive modification of the pVQD algorithm, that we indicate as pVQD with block extensions.
In this case, a new step of the Trotterized variational ansatz is added to the circuit once the optimization procedure does not reach the desired accuracy.
While this approach does improve the performance of the pVQD algorithm, we remark that it is not general, as it depends on the ansatz structure we have chosen.
Furthermore, we can see from the bottom panel of \cref{fig:xyz_floquet_statev} that the Adaptive pVQD method  always produces shallower circuits, with resources tailored to the needs of the specific time step.

\begin{figure}[h]
    \centering
    \includegraphics[width=\columnwidth]{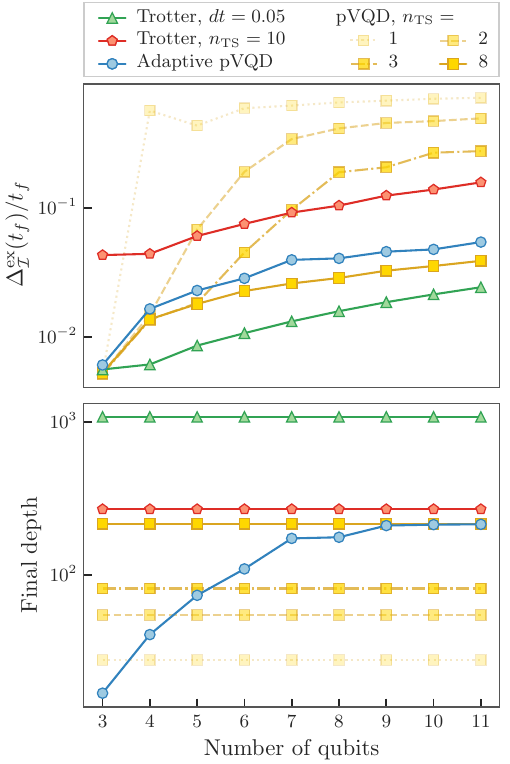}
    \caption{
    Adaptive pVQD algorithm with local pool compared to standard Trotter evolution and pVQD for the driven Heisenberg XYZ model.
    We employ the same setup indicated in \cref{fig:xyz_floquet_statev} for multiple systems of size $L \in \left[ 3,11 \right]$.
    The top panel shows the integrated exact infidelity of pVQD and Trotterization over an entire time evolution with final time $t_f=2$ as a function of the system size.
    The bottom panel shows the circuit depth at the end of the time evolution.
    }
    \label{fig:xyz_floquet_infid_results}
\end{figure}

\begin{figure}[h]
    \centering
    \includegraphics[width=\columnwidth]{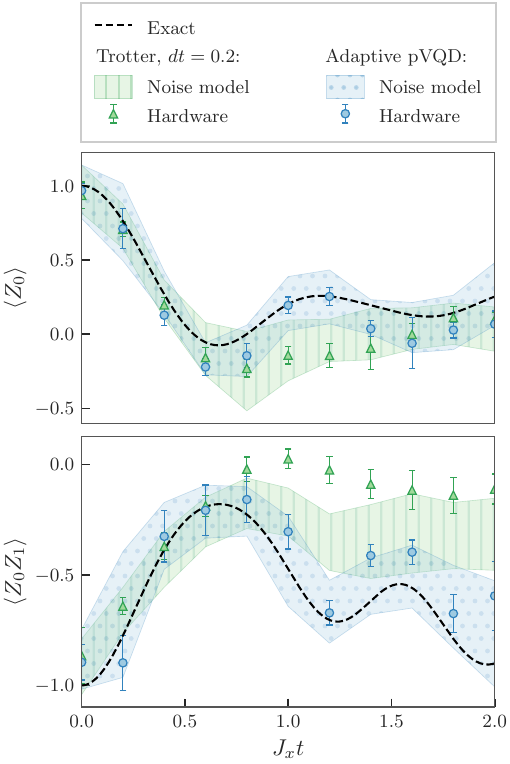}
    \caption{
    Observables measured with the IBM Manila device for the driven Heisenberg XYZ model on an open chain with 4 sites, $J_x=1, J_y=0.8, J_z=0.6$ and an antiferromagnetic initial state $|\psi_0\rangle = |0101\rangle $.
    The Trotter simulation is performed with a fixed Trotter step size of $dt=0.2$.
    The Adaptive pVQD circuits $|\psi(\bm{\theta})\rangle$ were obtained with a noiseless simulation that used a local operator pool.
    The shaded areas correspond to 50 noisy simulations using the noise model of IBM Manila. 
    Each data point and error bar correspond to the mean and the standard deviation, respectively, of 50 experiments performed on hardware.
    Zero noise extrapolation was applied to both noisy simulations and hardware experiments. 
    Idle qubits were also dynamically decoupled from the active ones.
    }
\label{fig:xyz_floquet_results_on_hardware}
\end{figure}

Then, we extend the study to systems with different sizes.
To this end, we define the integrated exact infidelity 

\begin{equation}
\label{eq:integrated_infidelity}
    \Delta_\mathcal{I}^\mathrm{ex}(t_f) = \int_{0}^{t_f} \left( 1 - |\langle \Psi(t) | \psi(\bm{\theta}) \rangle|^{2} \right) dt 
\end{equation}

with respect to the exact wave function $\ket{\Psi(t)}$ computed on a classical device.
We again fix a final evolution time $t_f = 2$ and evaluate $\Delta_\mathcal{I}^\mathrm{ex}(t_f)$ for each method for systems of $ L \in \left[3,11\right] $ spins.
In particular, we consider a Trotter circuit with a fixed depth of $n_\mathrm{TS}=10$ and one with fixed Trotter step size $dt=J_x t/n_\mathrm{TS}=0.05$, the same we use in the Trotter step of the pVQD algorithm.
The results are shown in \cref{fig:xyz_floquet_infid_results}, together with the circuit depth at the end of the time evolution.

We note that the depth of the Adaptive pVQD circuits increases with the system size and converges to the Trotter circuit with fixed depth, while having a lower integrated exact infidelity.
We highlight that \cref{fig:xyz_floquet_infid_results} only indicates the depth of the final circuit.
In the case of Adaptive pVQD, this corresponds to the deepest circuit prepared.
The Trotterized circuits with a fixed Trotter step size yield the lowest values for $\Delta_\mathcal{I}^\mathrm{ex}$, but $n_\mathrm{TS} = 40$ Trotter steps are required to evolve the system to $t_f = 2$, resulting in circuits almost one order of magnitude deeper than any other.
We performed multiple pVQD simulations with different variational ansätze equivalent to $n_\mathrm{TS} = 1,2,3,8$ Trotter steps.
We note that the integrated exact infidelities of pVQD with $n_\mathrm{TS} = 1,2,3$ have a steep transition when the number of gates becomes smaller than the adaptive circuit.
This phenomenon suggests that the ansatz limitation is the main source of error in the variational calculations, while the adaptive circuit is able to increase effectively its representation power.
On the other hand, the standard pVQD calculation with $n_\mathrm{TS} = 8$ never undergoes this transition.
While the integrated exact infidelity is always lower than the adaptive approach, we have to note that the entire time evolution is performed with a deeper circuit.
Finally, we note a plateau in the depth of the circuit required by the adaptive algorithm when $L>8$.
This is similar to what observed in \cite{2021_yao_adaptive_tdva}, where the system size at which the number of gates required saturates depends on the evolution time.

The adaptive method is able to produce circuits that are orders of magnitude shallower than Trotterization while keeping the accuracy comparable to it.
Those circuit can be used to improve the measurement of observables at long times on current quantum devices, which are otherwise limited by the depth of the Trotterization.
For this reason, we first run the Adaptive pVQD algorithm on the simulator and use the resulting sets of variational parameters to prepare quantum circuit on the hardware for a system of $L=4$ spins.
In \cref{fig:xyz_floquet_results_on_hardware}, we compare observables measured both on those variational wave functions and on Trotterized circuits with a fixed Trotter step size of $dt=0.2$.

In this experiment, the final Trotter circuit has 180 CNOTs.
This circuit is beyond what is currently accessible on quantum devices, settling the expectation value of the correlator close to $0$ for $J_x t > 0.8$.
On the other hand, the Adaptive pVQD parameterized circuit $|\psi(\bm{\theta})\rangle$ has 28 CNOTs at the end of the evolution.
This improvement in the number of gates is crucial for the application of error mitigation techniques, especially at longer times.
In particular, zero noise extrapolation (ZNE \cite{Temme_2017_zne,2017_ying_vqa_example})  was applied both on the noisy simulations and hardware experiments.
We choose a quadratic fit on values obtained with noise scaling factors $\left[ 1,2,3 \right]$.
Moreover, when running our algorithm on hardware, we dynamically decouple the idle qubits from the active ones using the standard procedure available in Qiskit \cite{qiskit}.
We expect that more advanced noise mitigation techniques, such as the one presented in \cite{VandenBerg2023}, will improve the results on the Trotter circuit.
However, this is also true for the variational circuit prepared by the Adaptive pVQD.

\subsection{Fermi-Hubbard model}

The Hamiltonian of the Fermi-Hubbard model on a $L_x \times L_y$ rectangular lattice is given by
\begin{equation} \label{eq:hubbard_hamiltonian}
    H = -J \sum_{ \langle i j \rangle, \sigma } (c_{i \sigma}^\dagger c_{j \sigma} + c_{j \sigma}^\dagger c_{i \sigma}) + U \sum_{i=0}^{L_x L_y-1} n_{i \uparrow} n_{i \downarrow},
\end{equation}

where $c_{i \sigma}^\dagger$ ($c_{i \sigma}$) is the creation (annihilation) fermionic operator of spin $\sigma \in \{ \uparrow, \downarrow \}$ at site $i$, $n_{i \sigma} = c_{i \sigma}^\dagger c_{i \sigma}$ counts the number of fermions with spin $\sigma$ at site $i$ and $\langle i j \rangle$ denotes nearest neighbor sites on the lattice.
The first term in the Hamiltonian accounts for the hopping between nearest neighbor lattice sites, while the second term describes the on-site interactions.

There are several ways to encode fermionic Hamiltonians into qubit operators \cite{Jordan93,Bravyi02_bk,Verstraete_2005,Whitfield_2016,Setia_2019,Chen_2022,Nys_2023}.
In this work, we consider the Jordan-Wigner mapping \cite{Jordan93} to encode each fermionic mode into a qubit.
Since every lattice site can host two modes ($\uparrow$, $\downarrow$), $N = 2 L_x L_y$ qubits are required to simulate the Fermi-Hubbard model on a $L_x \times L_y$ grid. 
Before performing a fermionic encoding, we eliminate the spin index via $c_{i \uparrow} \to c_i$ and $c_{i \downarrow} \to c_{i + N/2}$ (and analogously for the number operator $n_{i \sigma}$). 
We then  map each fermionic operator into a spin operator:
\begin{align}
    c_i & \to Z^{\otimes i} \otimes \sigma^+ \otimes \mathbb{I}^{\otimes N-i-1}, \\
    c_i^\dagger & \to Z^{\otimes i} \otimes \sigma^- \otimes \mathbb{I}^{\otimes N-i-1},
\end{align}
where $\sigma^{\pm} = (X \pm i Y)/2$.
The local occupation number can then be identified with the local spin number according to $n_i \in \{ 0,1 \} \mapsto Z_i \in \{ \uparrow, \downarrow \}$.
More details on the fermionic indexing convention and implementing a Trotter step can be found in \cref{appendix:Trotter step circuits}.

\begin{figure}
    \centering
    \includegraphics[width=\columnwidth]{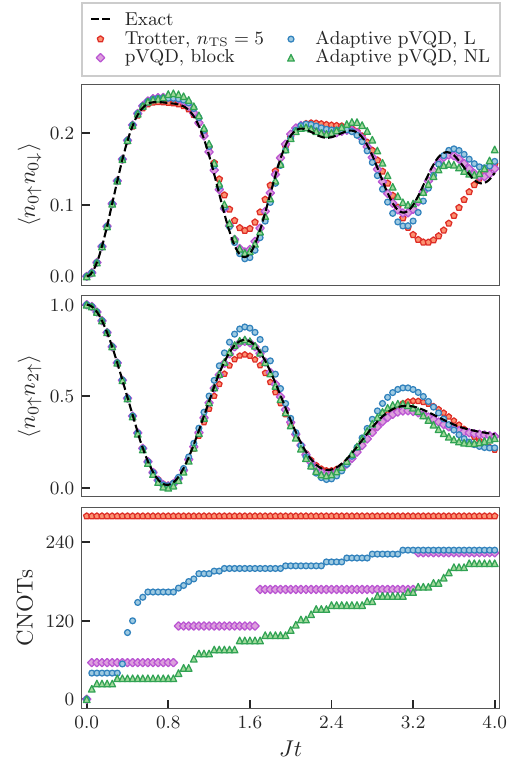}
    \caption{
    Adaptive pVQD schemes for the Fermi-Hubbard model on a $2 \times 2$ open square lattice (8 qubits) with $U/J = 0.8$.
    Local (L) and non-local (NL) operator pools are used to perform noiseless simulations and the results are compared to a Trotter evolution with $n_\mathrm{TS}=5$ Trotter steps and pVQD with block extensions.
    The system starts in the half-filled antiferromagnetic state $| \psi_0 \rangle = | n_{0 \uparrow} n_{1 \uparrow} n_{2 \uparrow} \cdots \rangle = |10100101 \rangle$.
    We fixed the infidelity threshold to $\varepsilon = 10^{-4}$.
    The top and middle panels show the expectation values of an on-site number density operator and a number density correlator over time.
    The bottom panel shows the number of CNOTs in the circuit describing the time-evolved wave function.
    }
    \label{fig:hubbard_2x2_results}
\end{figure}

Given that the mapping requires an ordering of the fermionic modes, operators that are local in space might generate very long Pauli strings.
For example, considering the snake-like pattern, vertical hopping terms generate strings of Pauli $Z$ with sizes up to $2L_x -2$.
This represents a bottleneck in studying fermionic systems with dimensionality higher than 1 on current quantum devices.
By restricting the operator pool, we investigate the possibility of describing time-evolved wave functions of the 2D Hubbard model using only local gates.
We perform noiseless simulations of a $2 \times 2$ square lattice, comparing local and non-local operator pools.
In particular, we measure the expectation values of a local density operator and a density correlator and count the number of CNOTs in the circuits. We use a fixed-depth Trotter simulation and a pVQD with block extension as a benchmark. 
The results are shown in \cref{fig:hubbard_2x2_results}.

We do not restrict ourselves to specific quantum hardware to keep the comparison as general as possible.
Instead, we count the number of CNOTs in a circuit by transpiling it into an abstract device with all-to-all connectivity that is able to perform arbitrary single qubit rotations and CNOTs.
The local and non-local pool variants show different behavior over time in the count of CNOTs. 
We note that the non-local variant always requires fewer CNOTs than its local counterpart.
However, some CNOTs are long-range, and their implementation on an actual device can be challenging on hardware with fixed topology and limited connectivity.
In contrast, the circuit structure produced by the local pool variant is already suited for current hardware implementation.
More details about the Adaptive pVQD output circuits can be found in \cref{appendix:output_circuits}.
Moreover, the plot highlights another limitation of the naive pVQD with block extensions approach.
Indeed, it always prepare more expensive circuits than the Adaptive pVQD with non local pool and in the end it has similar CNOT requirement to the local variant, while being restricted to use long range gates as required by the Trotter step.

\section{Conclusions}
\label{sec:conclusions}

We presented an adaptive version of pVQD, called Adaptive pVQD, to simulate the real-time evolution of quantum systems.
This algorithm importantly circumvents the need to choose a fixed ansatz from the beginning of the time evolution.
The parameterized quantum circuits are grown adaptively to be both problems and hardware-tailored.
This is obtained with a measurement overhead required to determine the best gate among those included in the operator pool.

However, the gate search can be operated in parallel and, in our scheme, does not involve circuits with auxiliary qubits.
This makes the Adaptive pVQD algorithm more hardware-efficient than standard methods, as exemplified in this work with the driven Heisenberg model on the IBM quantum hardware.
Finally, we have simulated the dynamics of the 2D Hubbard model with only local gates, using the adaptive procedure to mitigate one of the bottlenecks that current quantum devices face in studying fermionic systems.
Given the ease of introduction to the standard pVQD algorithm and its benefits, we believe that the adaptive procedure described here can be of great use in the simulation of dynamics both for current and future quantum devices.

\section*{Data availability}
The code used to run the simulations is open source and can be found at \cite{github}.
It was written in Python using Qiskit \cite{qiskit}.
Exact classical simulations were performed using Qutip \cite{qutip}.

\section*{Acknowledgments}
We thank S. Economou for insightful discussions.
This research was supported by the NCCR MARVEL, a National Centre of Competence in Research, funded by the Swiss National Science Foundation (grant number 205602).

\appendix

\section{Minimization routine} 
\label{appendix:Minimization routine}

Here we present additional details on the minimization routine that we applied throughout the simulations we presented in the main text.
In particular, we follow a gradient-based approach, with gradient computed using the parameter-shift rule.
Gradient-based and non-gradient-based optimization algorithms for dynamics were previously used for instance in \cite{2021_barison_p-vqd} and \cite{2022_berthusen_dynamics_on_hardware}, for both ideal and noisy quantum simulations.
 The parameter shift rule readily applies here since every Pauli string $A_i$ is involutory, i.e. $A_i^2 = \mathbb{I}$ \cite{2021_mari_param_shift_rule}. For a fixed set of operators $\bm{A}$, the gradient of the infidelity was thus computed via the parameter shift rule:
\begin{equation} \label{eq:parameter_shift_rule}
    \frac{\partial \mathcal{I}}{\partial d\theta_i} = \frac{ \mathcal{I}(\bm{\theta} + \bm{d} \theta + s \bm{e}_i) - \mathcal{I}(\bm{\theta} + \bm{d} \theta - s \bm{e}_i) }{ 2 \sin s },
\end{equation}
where $\bm{e}_i$ is the standard unit vector, and we fixed $s=\pi/2$.
The gradient was then fed to Adam \cite{2014_kingma_adam_optimizer}, implemented with the default hyperparameters and a learning rate $\alpha = 0.005$.
The shift parameters $\bm{d \theta}^*$  were consequently obtained using Adam.

Two stopping criteria for the optimizer were used: (1) the $\ell_\infty$-norm of the gradient of the infidelity is below a tolerance and (2) a maximum number of iterations is reached.
Fianlly, as showed in \cite{2021_barison_p-vqd}, an optimization threshold independent from $\Delta t$ can be used if $ \mathcal{I}$ is substituted with $ \mathcal{I} / \Delta t ^2$ as cost function.

\section{Gradient of the Fidelity} 
\label{appendix:gradient_fidelity}
In this Appendix, we derive the expression for the gradient of the adaptive step presented in \cref{eq:grad_fidelity}.
Given the quantum circuit $U(\bm{\theta})$ that prepares the state $| \psi (\bm{\theta}) = U(\bm{\theta}) | \psi_0 \rangle$, we want to add the gate $e^{-i d\theta_a A_a}$ to it, defining the new state $| \psi (\bm{\theta}+\bm{d \theta}) \rangle = U(\bm{\theta}) \, e^{-i d\theta_a A_a} | \psi_0 \rangle$.
To obtain the gradient of the fidelity with respect to this added parameter $d \theta_a$, it is convenient to first rewrite the fidelity given in \cref{eq:fidelity} as follows
\begin{align} 
    \mathcal{F}(\bm{d \theta},\Delta t) 
    &= | \langle \psi( \bm{\theta}+\bm{d \theta}) | U_\text{TS} (\Delta t) | \psi  ( \bm{\theta} ) \rangle |^2 \nonumber \\
    &= | \langle \psi_0 | e^{i d\theta_a A_a} U^\dagger(\bm{\theta}) U_\text{TS}(\Delta t) U(\bm{\theta}) | \psi_0 \rangle |^2 \nonumber \\
    &= \langle \psi_0 | e^{i d\theta_a A_a} U^\dagger(\bm{\theta}) U_\text{TS}(\Delta t) U(\bm{\theta}) | \psi_0 \rangle \nonumber\\
    & \quad * \langle \psi_0 | U^\dagger(\bm{\theta}) U_\text{TS}^\dagger(\Delta t) U(\bm{\theta}) e^{-i d\theta_a A_a} | \psi_0 \rangle \nonumber\\
    &= \langle \psi_0 | U^\dagger(\bm{\theta}) U_\text{TS}^\dagger(\Delta t) U(\bm{\theta}) e^{-i d\theta_a A_a} | \psi_0 \rangle \nonumber\\
    & \quad * \langle \psi_0 | e^{i d\theta_a A_a} U^\dagger(\bm{\theta}) U_\text{TS}(\Delta t) U(\bm{\theta}) | \psi_0 \rangle \nonumber\\
    &= \langle \phi(\bm{\theta}, \Delta t) | e^{-i d\theta_a A_a} P_0 e^{i d\theta_a A_a} | \phi(\bm{\theta}, \Delta t) \rangle,
\end{align}
where we defined $| \phi(\bm{\theta}, \Delta t) \rangle = U^\dagger(\bm{\theta}) U_\text{TS}(\Delta t) U(\bm{\theta}) | \psi_0 \rangle$ and the projector $P_0 = | \psi_0 \rangle \langle \psi_0 |$.
One can then readily differentiate with respect to $d \theta_a$ to obtain
\begin{align}
    \frac{ \partial \mathcal{F} }{ \partial d\theta_a } 
    &= \bra{\phi(\bm{\theta},\Delta t )}  e^{-id\theta_a A_a} [P_0,iA_a] e^{id\theta_a A_a} \ket{\phi(\bm{\theta},\Delta t )}
\end{align}
which precisely corresponds to \cref{eq:grad_fidelity}.

\section{Adaptive step implementation}
\label{appendix:tetris_adaptive_steps}

In this Appendix we illustrate the adaptive procedure we have used in our simulations, based on what was initially proposed in \cite{2022_economou_tetris}.
The overall procedure can be divided in the following steps:

\begin{enumerate}

\item \label{item:calculate_gradient} \textit{Compute the gradient of the fidelity for each operator in the pool}.
To process the pool, the gate $e^{-i \theta_a A_a}$ associated to each trial operator $A_a \in \mathcal{A}$ is appended one at a time to the current parameterized circuit $\{ \bm{\theta}, \bm{A} \}$, resulting in the trial circuit $\{ (\bm{\theta}, 0), (\bm{A}, A_a ) \}$.
For the trajectory in parameter space to remain continuous, the new parameter $\theta_a$ is set to 0. 
The gradient of the fidelity with respect to the new parameter is computed for each trial circuit using the parameter shift rule, given explicitly in \cref{eq:parameter_shift_rule}.

\item \label{item:choose_gate_max_grad} \textit{Pick the operator in the pool that maximizes the gradient}. Update the parameters and operators to $\bm{\theta} \to (\bm{\theta},0)$ and $\bm{A} \to (\bm{A}, A^*)$, where $A^*$ is the operator $A_a$ that maximizes the fidelity gradient. 

\item \textit{Remove the operators in the pool that act on qubit(s) already acted on}.
Given that the operator $A^*$ obtained in \cref{item:choose_gate_max_grad} acts on the qubits indices $\bm{\alpha}$, the subset of the operator pool that also acts on at least one index in $\bm{\alpha}$, namely
\begin{equation}
    \mathcal{A}_{\bm{\alpha}} = \{ A_a | A_a \in \mathcal{A} \ \text{acts on} \ \bm{\beta}, \bm{\beta} \cup \bm{\alpha} \ne \emptyset  \}
\end{equation}
should be removed from the current operator pool.
Hence the pool can be updated as follows: $\mathcal{A} \to \mathcal{A} \setminus \mathcal{A}_{\bm{\alpha}}$.

\item \textit{Go back to \cref{item:choose_gate_max_grad} until the operator pool is empty}.

\item \label{item:return_step} \textit{Return the new circuit}. The new parameterized circuit is characterized by $\bm{\theta} \to (\bm{\theta}, 0, \cdots, 0)$ and $\bm{A} \to (\bm{A}, A_0^*, A_1^*, \cdots, A_k^*)$, assuming that $k$ new operators were added.

\end{enumerate} 

As stated in the main text, this procedure guarantees that the depth of the parameterized circuit $|\psi(\bm{\theta})\rangle$ is increased by 1 in each adaptive step \cite{2022_economou_tetris}.

\section{Adaptive pVQD output circuits} 
\label{appendix:output_circuits}

\begin{figure}[h]
    \centering
    \includegraphics[width=\columnwidth]{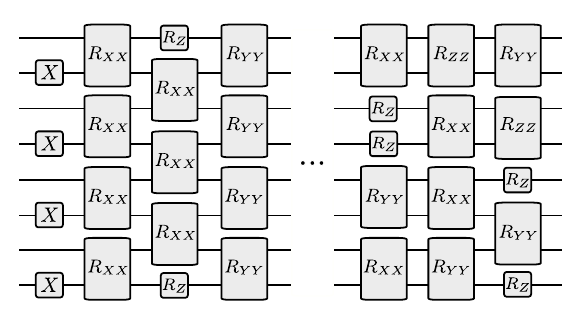}
    \caption{
    Variational circuit obtained at $J_x t=2$  in the simulation shown in \cref{fig:xyz_floquet_infid_results}, using the Adaptive pVQD algorithm and local operator pool.
    }
    \label{fig:final_circ_xyz}
\end{figure}

\begin{figure}[h]
    \centering
    \includegraphics[width=\columnwidth]{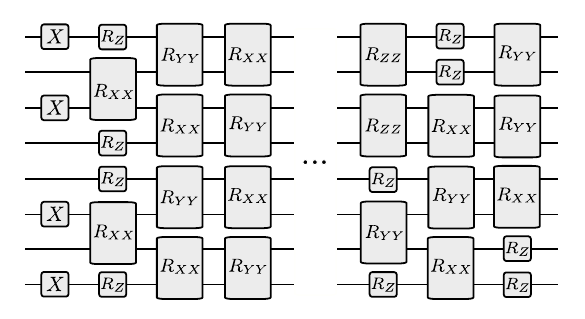}
    \caption{
     Variational circuit obtained at $J t=4$  in the simulation shown in \cref{fig:hubbard_2x2_results}, using the Adaptive pVQD algorithm and local operator pool.
    }
    \label{fig:final_circ_hubbard}
\end{figure}

We illustrate in \cref{fig:final_circ_xyz,fig:final_circ_hubbard} examples of parameterized circuits obtained with the Adaptive pVQD algorithm in simulations shown in the main text. 
Each column of operators in the circuits corresponds to an adaptive step.

\section{Trotter step circuit encodings} 
\label{appendix:Trotter step circuits}

\begin{figure}[h]
    \centering
    \includegraphics[width=\columnwidth]{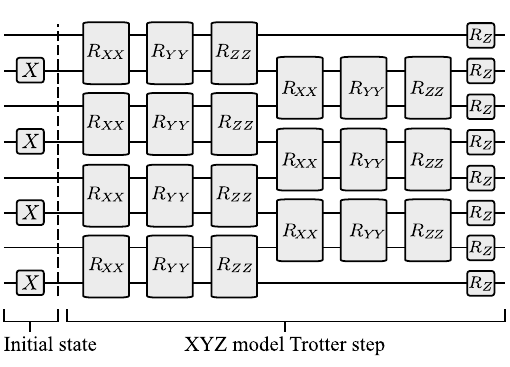}
    \caption{Implementation of an antiferromagnetic initial state and a Trotter step for the driven Heisenberg model given in \cref{eq:driven_heisenberg_model}.}
    \label{fig:driven_xyz_trotter_step}
\end{figure}

In this Appendix we provide the circuits we used to implement a single Trotter step of the driven Heisenberg and the Hubbard models.
The Trotter step in the driven Heisenberg model is implemented with a checkerboard pattern of the two qubit gates $R_{XX}, R_{YY}, R_{ZZ}$, with a layer of single qubit $R_{Z}$ at the end.
We show a sketch  in \cref{fig:driven_xyz_trotter_step}. 

\begin{figure}[h]
    \centering
    \includegraphics[width=\columnwidth]{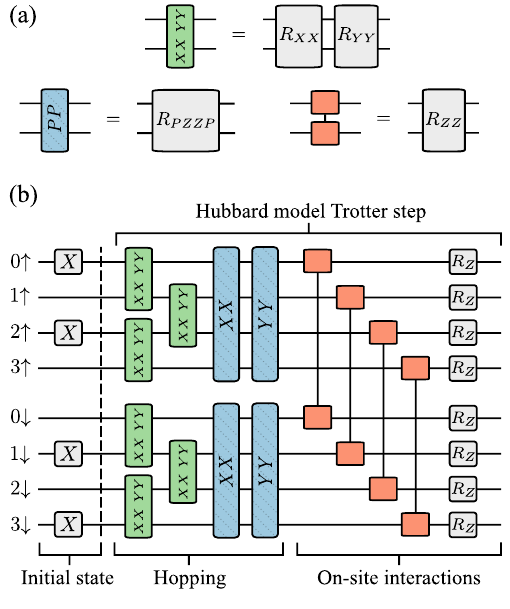}
    \caption{
    (a) Gates used to define a Trotter step. 
    (b) Quantum circuit encoding the first order Trotter step of the Hubbard model with an half-filled antiferromagnetic initial state.
    }
    \label{fig:hubbard_trotter_step}
\end{figure}

To realize the Trotter circuit for the Hubbard model, we first have to establish an ordering in the latices sites and the modes.
We number the sites using a snake-like pattern and, as indicated in the main text, we eliminate the spin index via $c_{i \uparrow} \to c_i$ and $c_{i \downarrow} \to c_{i + N/2}$.
Under this ordering, the Jordan-Wigner transformation of the Hamiltonian terms reads

\begin{align}
    c_{i\uparrow}^\dagger c_{j \uparrow} + c_{j \uparrow}^\dagger c_{i \uparrow} &\mapsto \frac{1}{2} \left[ X_i \prod_{k=i+1}^{j-1} Z_k   X_{j} + Y_i  \prod_{k=i+1}^{j-1} Z_k Y_{j} \right], \\
    c_{i\downarrow}^\dagger c_{j \downarrow} + c_{j \downarrow}^\dagger c_{i \downarrow} &\mapsto \frac{1}{2} \Bigg[ X_{i+N/2} \prod_{k=i+1}^{j-1} Z_{k+N/2}   X_{j+N/2} \, +  \\
    &\, \,\,\,\,\,\,\,\,\,\, + Y_{i+N/2}  \prod_{k=i+1}^{j-1} Z_{k+N/2} Y_{j+N/2} \Bigg], \nonumber \\ 
    n_{i \uparrow} n_{i \downarrow} &\mapsto \frac{1}{4} (\mathbb{I}-Z_i)(\mathbb{I}-Z_{i+N/2}),
\end{align}

where we assumed $j>i$ without loss of generality.
Given the mapped Hamiltonian, the Trotter step can not be implemented using only $R_{XX}, R_{YY}, R_{ZZ}$ and $R_Z$ gates.
Indeed, the non locality of the mapping requires some multi-qubit rotation with size up to $2L_{x}$.
The two multi-qubit gates are the rotations generated by the Pauli strings $XZZX$ and $YZZY$, which can be decomposed as shown in \cite{Tacchino_2019}.
\cref{fig:hubbard_trotter_step} presents our implementation.

\clearpage
\newpage
\twocolumngrid


\begin{thebibliography}{66}%
\makeatletter
\providecommand \@ifxundefined [1]{%
 \@ifx{#1\undefined}
}%
\providecommand \@ifnum [1]{%
 \ifnum #1\expandafter \@firstoftwo
 \else \expandafter \@secondoftwo
 \fi
}%
\providecommand \@ifx [1]{%
 \ifx #1\expandafter \@firstoftwo
 \else \expandafter \@secondoftwo
 \fi
}%
\providecommand \natexlab [1]{#1}%
\providecommand \enquote  [1]{``#1''}%
\providecommand \bibnamefont  [1]{#1}%
\providecommand \bibfnamefont [1]{#1}%
\providecommand \citenamefont [1]{#1}%
\providecommand \href@noop [0]{\@secondoftwo}%
\providecommand \href [0]{\begingroup \@sanitize@url \@href}%
\providecommand \@href[1]{\@@startlink{#1}\@@href}%
\providecommand \@@href[1]{\endgroup#1\@@endlink}%
\providecommand \@sanitize@url [0]{\catcode `\\12\catcode `\$12\catcode
  `\&12\catcode `\#12\catcode `\^12\catcode `\_12\catcode `\%12\relax}%
\providecommand \@@startlink[1]{}%
\providecommand \@@endlink[0]{}%
\providecommand \url  [0]{\begingroup\@sanitize@url \@url }%
\providecommand \@url [1]{\endgroup\@href {#1}{\urlprefix }}%
\providecommand \urlprefix  [0]{URL }%
\providecommand \Eprint [0]{\href }%
\providecommand \doibase [0]{https://doi.org/}%
\providecommand \selectlanguage [0]{\@gobble}%
\providecommand \bibinfo  [0]{\@secondoftwo}%
\providecommand \bibfield  [0]{\@secondoftwo}%
\providecommand \translation [1]{[#1]}%
\providecommand \BibitemOpen [0]{}%
\providecommand \bibitemStop [0]{}%
\providecommand \bibitemNoStop [0]{.\EOS\space}%
\providecommand \EOS [0]{\spacefactor3000\relax}%
\providecommand \BibitemShut  [1]{\csname bibitem#1\endcsname}%
\let\auto@bib@innerbib\@empty
\bibitem [{\citenamefont {Sandvik}\ \emph {et~al.}(2010)\citenamefont
  {Sandvik}, \citenamefont {Avella},\ and\ \citenamefont
  {Mancini}}]{2010_sandvik_computational_methods}%
  \BibitemOpen
  \bibfield  {author} {\bibinfo {author} {\bibfnamefont {A.~W.}\ \bibnamefont
  {Sandvik}}, \bibinfo {author} {\bibfnamefont {A.}~\bibnamefont {Avella}},\
  and\ \bibinfo {author} {\bibfnamefont {F.}~\bibnamefont {Mancini}},\
  }\bibfield  {title} {\bibinfo {title} {Computational studies of quantum spin
  systems},\ }in\ \href {https://doi.org/10.1063/1.3518900} {\emph {\bibinfo
  {booktitle} {{AIP} Conference Proceedings}}}\ (\bibinfo  {publisher}
  {{AIP}},\ \bibinfo {year} {2010})\BibitemShut {NoStop}%
\bibitem [{\citenamefont {Carleo}\ and\ \citenamefont
  {Troyer}(2017)}]{2017_carleo_nqs}%
  \BibitemOpen
  \bibfield  {author} {\bibinfo {author} {\bibfnamefont {G.}~\bibnamefont
  {Carleo}}\ and\ \bibinfo {author} {\bibfnamefont {M.}~\bibnamefont
  {Troyer}},\ }\bibfield  {title} {\bibinfo {title} {Solving the quantum
  many-body problem with artificial neural networks},\ }\href
  {https://doi.org/10.1126/science.aag2302} {\bibfield  {journal} {\bibinfo
  {journal} {Science}\ }\textbf {\bibinfo {volume} {355}},\ \bibinfo {pages}
  {602} (\bibinfo {year} {2017})}\BibitemShut {NoStop}%
\bibitem [{\citenamefont
  {Or{\'{u}}s}(2019)}]{2019_orus_tensor_networks_review}%
  \BibitemOpen
  \bibfield  {author} {\bibinfo {author} {\bibfnamefont {R.}~\bibnamefont
  {Or{\'{u}}s}},\ }\bibfield  {title} {\bibinfo {title} {Tensor networks for
  complex quantum systems},\ }\href {https://doi.org/10.1038/s42254-019-0086-7}
  {\bibfield  {journal} {\bibinfo  {journal} {Nature Reviews Physics}\ }\textbf
  {\bibinfo {volume} {1}},\ \bibinfo {pages} {538} (\bibinfo {year}
  {2019})}\BibitemShut {NoStop}%
\bibitem [{\citenamefont {Carleo}\ \emph {et~al.}(2019)\citenamefont {Carleo},
  \citenamefont {Cirac}, \citenamefont {Cranmer}, \citenamefont {Daudet},
  \citenamefont {Schuld}, \citenamefont {Tishby}, \citenamefont
  {Vogt-Maranto},\ and\ \citenamefont {Zdeborov{\'{a}
  }}}]{2019_carleo_ml_in_physics}%
  \BibitemOpen
  \bibfield  {author} {\bibinfo {author} {\bibfnamefont {G.}~\bibnamefont
  {Carleo}}, \bibinfo {author} {\bibfnamefont {I.}~\bibnamefont {Cirac}},
  \bibinfo {author} {\bibfnamefont {K.}~\bibnamefont {Cranmer}}, \bibinfo
  {author} {\bibfnamefont {L.}~\bibnamefont {Daudet}}, \bibinfo {author}
  {\bibfnamefont {M.}~\bibnamefont {Schuld}}, \bibinfo {author} {\bibfnamefont
  {N.}~\bibnamefont {Tishby}}, \bibinfo {author} {\bibfnamefont
  {L.}~\bibnamefont {Vogt-Maranto}},\ and\ \bibinfo {author} {\bibfnamefont
  {L.}~\bibnamefont {Zdeborov{\'{a} }}},\ }\bibfield  {title} {\bibinfo {title}
  {Machine learning and the physical sciences},\ }\bibfield  {journal}
  {\bibinfo  {journal} {Reviews of Modern Physics}\ }\textbf {\bibinfo {volume}
  {91}},\ \href {https://doi.org/10.1103/revmodphys.91.045002}
  {10.1103/revmodphys.91.045002} (\bibinfo {year} {2019})\BibitemShut {NoStop}%
\bibitem [{\citenamefont {Cady}\ \emph {et~al.}(2008)\citenamefont {Cady},
  \citenamefont {Crabtree},\ and\ \citenamefont {Brudvig}}]{Cady2008}%
  \BibitemOpen
  \bibfield  {author} {\bibinfo {author} {\bibfnamefont {C.~W.}\ \bibnamefont
  {Cady}}, \bibinfo {author} {\bibfnamefont {R.~H.}\ \bibnamefont {Crabtree}},\
  and\ \bibinfo {author} {\bibfnamefont {G.~W.}\ \bibnamefont {Brudvig}},\
  }\bibfield  {title} {\bibinfo {title} {Functional models for the
  oxygen-evolving complex of photosystem ii},\ }\href
  {https://doi.org/https://doi.org/10.1016/j.ccr.2007.06.002} {\bibfield
  {journal} {\bibinfo  {journal} {Coordination Chemistry Reviews}\ }\textbf
  {\bibinfo {volume} {252}},\ \bibinfo {pages} {444} (\bibinfo {year}
  {2008})},\ \bibinfo {note} {the Role of Manganese in Photosystem
  II}\BibitemShut {NoStop}%
\bibitem [{\citenamefont {Schimka}\ \emph {et~al.}(2010)\citenamefont
  {Schimka}, \citenamefont {Harl}, \citenamefont {Stroppa}, \citenamefont
  {Gr{\"{u}}neis}, \citenamefont {Marsman}, \citenamefont {Mittendorfer},\ and\
  \citenamefont {Kresse}}]{Schimka2010}%
  \BibitemOpen
  \bibfield  {author} {\bibinfo {author} {\bibfnamefont {L.}~\bibnamefont
  {Schimka}}, \bibinfo {author} {\bibfnamefont {J.}~\bibnamefont {Harl}},
  \bibinfo {author} {\bibfnamefont {A.}~\bibnamefont {Stroppa}}, \bibinfo
  {author} {\bibfnamefont {A.}~\bibnamefont {Gr{\"{u}}neis}}, \bibinfo {author}
  {\bibfnamefont {M.}~\bibnamefont {Marsman}}, \bibinfo {author} {\bibfnamefont
  {F.}~\bibnamefont {Mittendorfer}},\ and\ \bibinfo {author} {\bibfnamefont
  {G.}~\bibnamefont {Kresse}},\ }\bibfield  {title} {\bibinfo {title}
  {{Accurate surface and adsorption energies from many-body perturbation
  theory}},\ }\href {https://doi.org/10.1038/nmat2806} {\bibfield  {journal}
  {\bibinfo  {journal} {Nature Materials}\ }\textbf {\bibinfo {volume} {9}},\
  \bibinfo {pages} {741} (\bibinfo {year} {2010})}\BibitemShut {NoStop}%
\bibitem [{\citenamefont {Leggett}(2006)}]{Leggett2006}%
  \BibitemOpen
  \bibfield  {author} {\bibinfo {author} {\bibfnamefont {A.~J.}\ \bibnamefont
  {Leggett}},\ }\bibfield  {title} {\bibinfo {title} {{What DO we know about
  high Tc?}},\ }\href {https://doi.org/10.1038/nphys254} {\bibfield  {journal}
  {\bibinfo  {journal} {Nature Physics}\ }\textbf {\bibinfo {volume} {2}},\
  \bibinfo {pages} {134} (\bibinfo {year} {2006})}\BibitemShut {NoStop}%
\bibitem [{\citenamefont {Balents}(2010)}]{Balents2010}%
  \BibitemOpen
  \bibfield  {author} {\bibinfo {author} {\bibfnamefont {L.}~\bibnamefont
  {Balents}},\ }\bibfield  {title} {\bibinfo {title} {{Spin liquids in
  frustrated magnets}},\ }\href {https://doi.org/10.1038/nature08917}
  {\bibfield  {journal} {\bibinfo  {journal} {Nature}\ }\textbf {\bibinfo
  {volume} {464}},\ \bibinfo {pages} {199} (\bibinfo {year}
  {2010})}\BibitemShut {NoStop}%
\bibitem [{\citenamefont {Arute}\ \emph {et~al.}(2019)\citenamefont {Arute}
  \emph {et~al.}}]{2019_arute_quantum_supremacy}%
  \BibitemOpen
  \bibfield  {author} {\bibinfo {author} {\bibfnamefont {F.}~\bibnamefont
  {Arute}} \emph {et~al.},\ }\bibfield  {title} {\bibinfo {title} {Quantum
  supremacy using a programmable superconducting processor},\ }\href
  {https://doi.org/10.1038/s41586-019-1666-5} {\bibfield  {journal} {\bibinfo
  {journal} {Nature}\ }\textbf {\bibinfo {volume} {574}},\ \bibinfo {pages}
  {505} (\bibinfo {year} {2019})}\BibitemShut {NoStop}%
\bibitem [{\citenamefont {Zhong}\ \emph {et~al.}(2020)\citenamefont {Zhong}
  \emph {et~al.}}]{2020_zhong_quantum_advantage_photons}%
  \BibitemOpen
  \bibfield  {author} {\bibinfo {author} {\bibfnamefont {H.-S.}\ \bibnamefont
  {Zhong}} \emph {et~al.},\ }\bibfield  {title} {\bibinfo {title} {Quantum
  computational advantage using photons},\ }\href
  {https://doi.org/10.1126/science.abe8770} {\bibfield  {journal} {\bibinfo
  {journal} {Science}\ }\textbf {\bibinfo {volume} {370}},\ \bibinfo {pages}
  {1460} (\bibinfo {year} {2020})}\BibitemShut {NoStop}%
\bibitem [{\citenamefont {Huang}\ \emph {et~al.}(2022)\citenamefont {Huang},
  \citenamefont {Broughton}, \citenamefont {Cotler}, \citenamefont {Chen},
  \citenamefont {Li}, \citenamefont {Mohseni}, \citenamefont {Neven},
  \citenamefont {Babbush}, \citenamefont {Kueng}, \citenamefont {Preskill},\
  and\ \citenamefont {McClean}}]{2022_huang_quantum_advantage}%
  \BibitemOpen
  \bibfield  {author} {\bibinfo {author} {\bibfnamefont {H.-Y.}\ \bibnamefont
  {Huang}}, \bibinfo {author} {\bibfnamefont {M.}~\bibnamefont {Broughton}},
  \bibinfo {author} {\bibfnamefont {J.}~\bibnamefont {Cotler}}, \bibinfo
  {author} {\bibfnamefont {S.}~\bibnamefont {Chen}}, \bibinfo {author}
  {\bibfnamefont {J.}~\bibnamefont {Li}}, \bibinfo {author} {\bibfnamefont
  {M.}~\bibnamefont {Mohseni}}, \bibinfo {author} {\bibfnamefont
  {H.}~\bibnamefont {Neven}}, \bibinfo {author} {\bibfnamefont
  {R.}~\bibnamefont {Babbush}}, \bibinfo {author} {\bibfnamefont
  {R.}~\bibnamefont {Kueng}}, \bibinfo {author} {\bibfnamefont
  {J.}~\bibnamefont {Preskill}},\ and\ \bibinfo {author} {\bibfnamefont
  {J.~R.}\ \bibnamefont {McClean}},\ }\bibfield  {title} {\bibinfo {title}
  {Quantum advantage in learning from experiments},\ }\href
  {https://doi.org/10.1126/science.abn7293} {\bibfield  {journal} {\bibinfo
  {journal} {Science}\ }\textbf {\bibinfo {volume} {376}},\ \bibinfo {pages}
  {1182} (\bibinfo {year} {2022})}\BibitemShut {NoStop}%
\bibitem [{\citenamefont {Dolde}\ \emph {et~al.}(2014)\citenamefont {Dolde},
  \citenamefont {Bergholm}, \citenamefont {Wang}, \citenamefont {Jakobi},
  \citenamefont {Naydenov}, \citenamefont {Pezzagna}, \citenamefont {Meijer},
  \citenamefont {Jelezko}, \citenamefont {Neumann}, \citenamefont
  {Schulte-Herbrüggen}, \citenamefont {Biamonte},\ and\ \citenamefont
  {Wrachtrup}}]{2014_dolde_optimal_control}%
  \BibitemOpen
  \bibfield  {author} {\bibinfo {author} {\bibfnamefont {F.}~\bibnamefont
  {Dolde}}, \bibinfo {author} {\bibfnamefont {V.}~\bibnamefont {Bergholm}},
  \bibinfo {author} {\bibfnamefont {Y.}~\bibnamefont {Wang}}, \bibinfo {author}
  {\bibfnamefont {I.}~\bibnamefont {Jakobi}}, \bibinfo {author} {\bibfnamefont
  {B.}~\bibnamefont {Naydenov}}, \bibinfo {author} {\bibfnamefont
  {S.}~\bibnamefont {Pezzagna}}, \bibinfo {author} {\bibfnamefont
  {J.}~\bibnamefont {Meijer}}, \bibinfo {author} {\bibfnamefont
  {F.}~\bibnamefont {Jelezko}}, \bibinfo {author} {\bibfnamefont
  {P.}~\bibnamefont {Neumann}}, \bibinfo {author} {\bibfnamefont
  {T.}~\bibnamefont {Schulte-Herbrüggen}}, \bibinfo {author} {\bibfnamefont
  {J.}~\bibnamefont {Biamonte}},\ and\ \bibinfo {author} {\bibfnamefont
  {J.}~\bibnamefont {Wrachtrup}},\ }\bibfield  {title} {\bibinfo {title}
  {High-fidelity spin entanglement using optimal control},\ }\bibfield
  {journal} {\bibinfo  {journal} {Nature Communications}\ }\textbf {\bibinfo
  {volume} {5}},\ \href {https://doi.org/10.1038/ncomms4371}
  {10.1038/ncomms4371} (\bibinfo {year} {2014})\BibitemShut {NoStop}%
\bibitem [{\citenamefont {Waldherr}\ \emph {et~al.}(2014)\citenamefont
  {Waldherr}, \citenamefont {Wang}, \citenamefont {Zaiser}, \citenamefont
  {Jamali}, \citenamefont {Schulte-Herbrüggen}, \citenamefont {Abe},
  \citenamefont {Ohshima}, \citenamefont {Isoya}, \citenamefont {Du},
  \citenamefont {Neumann},\ and\ \citenamefont
  {Wrachtrup}}]{2014_waldherr_optimal_control}%
  \BibitemOpen
  \bibfield  {author} {\bibinfo {author} {\bibfnamefont {G.}~\bibnamefont
  {Waldherr}}, \bibinfo {author} {\bibfnamefont {Y.}~\bibnamefont {Wang}},
  \bibinfo {author} {\bibfnamefont {S.}~\bibnamefont {Zaiser}}, \bibinfo
  {author} {\bibfnamefont {M.}~\bibnamefont {Jamali}}, \bibinfo {author}
  {\bibfnamefont {T.}~\bibnamefont {Schulte-Herbrüggen}}, \bibinfo {author}
  {\bibfnamefont {H.}~\bibnamefont {Abe}}, \bibinfo {author} {\bibfnamefont
  {T.}~\bibnamefont {Ohshima}}, \bibinfo {author} {\bibfnamefont
  {J.}~\bibnamefont {Isoya}}, \bibinfo {author} {\bibfnamefont {J.~F.}\
  \bibnamefont {Du}}, \bibinfo {author} {\bibfnamefont {P.}~\bibnamefont
  {Neumann}},\ and\ \bibinfo {author} {\bibfnamefont {J.}~\bibnamefont
  {Wrachtrup}},\ }\bibfield  {title} {\bibinfo {title} {Quantum error
  correction in a solid-state hybrid spin register},\ }\href
  {https://doi.org/10.1038/nature12919} {\bibfield  {journal} {\bibinfo
  {journal} {Nature}\ }\textbf {\bibinfo {volume} {506}},\ \bibinfo {pages}
  {204} (\bibinfo {year} {2014})}\BibitemShut {NoStop}%
\bibitem [{\citenamefont {Nam}\ \emph {et~al.}(2019)\citenamefont {Nam} \emph
  {et~al.}}]{2019_nam_ionq_progress}%
  \BibitemOpen
  \bibfield  {author} {\bibinfo {author} {\bibfnamefont {Y.}~\bibnamefont
  {Nam}} \emph {et~al.},\ }\href {https://doi.org/10.48550/ARXIV.1902.10171}
  {\bibinfo {title} {Ground-state energy estimation of the water molecule on a
  trapped ion quantum computer}} (\bibinfo {year} {2019})\BibitemShut {NoStop}%
\bibitem [{\citenamefont {Wan}\ \emph {et~al.}(2020)\citenamefont {Wan},
  \citenamefont {Jördens}, \citenamefont {Erickson}, \citenamefont {Wu},
  \citenamefont {Bowler}, \citenamefont {Tan}, \citenamefont {Hou},
  \citenamefont {Wineland}, \citenamefont {Wilson},\ and\ \citenamefont
  {Leibfried}}]{2020_wan_boulder_ion_traps}%
  \BibitemOpen
  \bibfield  {author} {\bibinfo {author} {\bibfnamefont {Y.}~\bibnamefont
  {Wan}}, \bibinfo {author} {\bibfnamefont {R.}~\bibnamefont {Jördens}},
  \bibinfo {author} {\bibfnamefont {S.~D.}\ \bibnamefont {Erickson}}, \bibinfo
  {author} {\bibfnamefont {J.~J.}\ \bibnamefont {Wu}}, \bibinfo {author}
  {\bibfnamefont {R.}~\bibnamefont {Bowler}}, \bibinfo {author} {\bibfnamefont
  {T.~R.}\ \bibnamefont {Tan}}, \bibinfo {author} {\bibfnamefont {P.-Y.}\
  \bibnamefont {Hou}}, \bibinfo {author} {\bibfnamefont {D.~J.}\ \bibnamefont
  {Wineland}}, \bibinfo {author} {\bibfnamefont {A.~C.}\ \bibnamefont
  {Wilson}},\ and\ \bibinfo {author} {\bibfnamefont {D.}~\bibnamefont
  {Leibfried}},\ }\bibfield  {title} {\bibinfo {title} {Ion transport and
  reordering in a 2d trap array},\ }\href
  {https://doi.org/10.1002/qute.202000028} {\bibfield  {journal} {\bibinfo
  {journal} {Advanced Quantum Technologies}\ }\textbf {\bibinfo {volume} {3}},\
  \bibinfo {pages} {2000028} (\bibinfo {year} {2020})}\BibitemShut {NoStop}%
\bibitem [{\citenamefont {Hughes}\ \emph {et~al.}(2020)\citenamefont {Hughes},
  \citenamefont {Sch\"afer}, \citenamefont {Thirumalai}, \citenamefont
  {Nadlinger}, \citenamefont {Woodrow}, \citenamefont {Lucas},\ and\
  \citenamefont {Ballance}}]{2020_hughes_high_fidelity_ent_gate}%
  \BibitemOpen
  \bibfield  {author} {\bibinfo {author} {\bibfnamefont {A.~C.}\ \bibnamefont
  {Hughes}}, \bibinfo {author} {\bibfnamefont {V.~M.}\ \bibnamefont
  {Sch\"afer}}, \bibinfo {author} {\bibfnamefont {K.}~\bibnamefont
  {Thirumalai}}, \bibinfo {author} {\bibfnamefont {D.~P.}\ \bibnamefont
  {Nadlinger}}, \bibinfo {author} {\bibfnamefont {S.~R.}\ \bibnamefont
  {Woodrow}}, \bibinfo {author} {\bibfnamefont {D.~M.}\ \bibnamefont {Lucas}},\
  and\ \bibinfo {author} {\bibfnamefont {C.~J.}\ \bibnamefont {Ballance}},\
  }\bibfield  {title} {\bibinfo {title} {Benchmarking a high-fidelity
  mixed-species entangling gate},\ }\href
  {https://doi.org/10.1103/PhysRevLett.125.080504} {\bibfield  {journal}
  {\bibinfo  {journal} {Phys. Rev. Lett.}\ }\textbf {\bibinfo {volume} {125}},\
  \bibinfo {pages} {080504} (\bibinfo {year} {2020})}\BibitemShut {NoStop}%
\bibitem [{Kim(2023)}]{Kim2023}%
  \BibitemOpen
  \bibfield  {title} {\bibinfo {title} {{Evidence for the utility of quantum
  computing before fault tolerance}},\ }\href
  {https://doi.org/10.1038/s41586-023-06096-3} {\bibfield  {journal} {\bibinfo
  {journal} {Nature}\ }\textbf {\bibinfo {volume} {618}},\ \bibinfo {pages}
  {500} (\bibinfo {year} {2023})}\BibitemShut {NoStop}%
\bibitem [{\citenamefont {Cerezo}\ \emph {et~al.}(2021)\citenamefont {Cerezo},
  \citenamefont {Arrasmith}, \citenamefont {Babbush}, \citenamefont {Benjamin},
  \citenamefont {Endo}, \citenamefont {Fujii}, \citenamefont {McClean},
  \citenamefont {Mitarai}, \citenamefont {Yuan}, \citenamefont {Cincio},\ and\
  \citenamefont {Coles}}]{2021_cerezo_vqa_rev}%
  \BibitemOpen
  \bibfield  {author} {\bibinfo {author} {\bibfnamefont {M.}~\bibnamefont
  {Cerezo}}, \bibinfo {author} {\bibfnamefont {A.}~\bibnamefont {Arrasmith}},
  \bibinfo {author} {\bibfnamefont {R.}~\bibnamefont {Babbush}}, \bibinfo
  {author} {\bibfnamefont {S.~C.}\ \bibnamefont {Benjamin}}, \bibinfo {author}
  {\bibfnamefont {S.}~\bibnamefont {Endo}}, \bibinfo {author} {\bibfnamefont
  {K.}~\bibnamefont {Fujii}}, \bibinfo {author} {\bibfnamefont {J.~R.}\
  \bibnamefont {McClean}}, \bibinfo {author} {\bibfnamefont {K.}~\bibnamefont
  {Mitarai}}, \bibinfo {author} {\bibfnamefont {X.}~\bibnamefont {Yuan}},
  \bibinfo {author} {\bibfnamefont {L.}~\bibnamefont {Cincio}},\ and\ \bibinfo
  {author} {\bibfnamefont {P.~J.}\ \bibnamefont {Coles}},\ }\bibfield  {title}
  {\bibinfo {title} {Variational quantum algorithms},\ }\href
  {https://doi.org/10.1038/s42254-021-00348-9} {\bibfield  {journal} {\bibinfo
  {journal} {Nature Reviews Physics}\ }\textbf {\bibinfo {volume} {3}},\
  \bibinfo {pages} {625} (\bibinfo {year} {2021})}\BibitemShut {NoStop}%
\bibitem [{\citenamefont {Bharti}\ \emph {et~al.}(2022)\citenamefont {Bharti},
  \citenamefont {Cervera-Lierta}, \citenamefont {Kyaw}, \citenamefont {Haug},
  \citenamefont {Alperin-Lea}, \citenamefont {Anand}, \citenamefont {Degroote},
  \citenamefont {Heimonen}, \citenamefont {Kottmann}, \citenamefont {Menke},
  \citenamefont {Mok}, \citenamefont {Sim}, \citenamefont {Kwek},\ and\
  \citenamefont {Aspuru-Guzik}}]{2022_bharti_nisq_algo_rev}%
  \BibitemOpen
  \bibfield  {author} {\bibinfo {author} {\bibfnamefont {K.}~\bibnamefont
  {Bharti}}, \bibinfo {author} {\bibfnamefont {A.}~\bibnamefont
  {Cervera-Lierta}}, \bibinfo {author} {\bibfnamefont {T.~H.}\ \bibnamefont
  {Kyaw}}, \bibinfo {author} {\bibfnamefont {T.}~\bibnamefont {Haug}}, \bibinfo
  {author} {\bibfnamefont {S.}~\bibnamefont {Alperin-Lea}}, \bibinfo {author}
  {\bibfnamefont {A.}~\bibnamefont {Anand}}, \bibinfo {author} {\bibfnamefont
  {M.}~\bibnamefont {Degroote}}, \bibinfo {author} {\bibfnamefont
  {H.}~\bibnamefont {Heimonen}}, \bibinfo {author} {\bibfnamefont {J.~S.}\
  \bibnamefont {Kottmann}}, \bibinfo {author} {\bibfnamefont {T.}~\bibnamefont
  {Menke}}, \bibinfo {author} {\bibfnamefont {W.-K.}\ \bibnamefont {Mok}},
  \bibinfo {author} {\bibfnamefont {S.}~\bibnamefont {Sim}}, \bibinfo {author}
  {\bibfnamefont {L.-C.}\ \bibnamefont {Kwek}},\ and\ \bibinfo {author}
  {\bibfnamefont {A.}~\bibnamefont {Aspuru-Guzik}},\ }\bibfield  {title}
  {\bibinfo {title} {Noisy intermediate-scale quantum algorithms},\ }\href
  {https://doi.org/10.1103/RevModPhys.94.015004} {\bibfield  {journal}
  {\bibinfo  {journal} {Rev. Mod. Phys.}\ }\textbf {\bibinfo {volume} {94}},\
  \bibinfo {pages} {015004} (\bibinfo {year} {2022})}\BibitemShut {NoStop}%
\bibitem [{\citenamefont {Yuan}\ \emph {et~al.}(2019)\citenamefont {Yuan},
  \citenamefont {Endo}, \citenamefont {Zhao}, \citenamefont {Li},\ and\
  \citenamefont {Benjamin}}]{2019_yuan_variational_simulations_rev}%
  \BibitemOpen
  \bibfield  {author} {\bibinfo {author} {\bibfnamefont {X.}~\bibnamefont
  {Yuan}}, \bibinfo {author} {\bibfnamefont {S.}~\bibnamefont {Endo}}, \bibinfo
  {author} {\bibfnamefont {Q.}~\bibnamefont {Zhao}}, \bibinfo {author}
  {\bibfnamefont {Y.}~\bibnamefont {Li}},\ and\ \bibinfo {author}
  {\bibfnamefont {S.~C.}\ \bibnamefont {Benjamin}},\ }\bibfield  {title}
  {\bibinfo {title} {Theory of variational quantum simulation},\ }\href
  {https://doi.org/10.22331/q-2019-10-07-191} {\bibfield  {journal} {\bibinfo
  {journal} {Quantum}\ }\textbf {\bibinfo {volume} {3}},\ \bibinfo {pages}
  {191} (\bibinfo {year} {2019})}\BibitemShut {NoStop}%
\bibitem [{\citenamefont {Peruzzo}\ \emph {et~al.}(2014)\citenamefont
  {Peruzzo}, \citenamefont {McClean}, \citenamefont {Shadbolt}, \citenamefont
  {Yung}, \citenamefont {Zhou}, \citenamefont {Love}, \citenamefont
  {Aspuru-Guzik},\ and\ \citenamefont {O'Brien}}]{2014_peruzzo_vqe}%
  \BibitemOpen
  \bibfield  {author} {\bibinfo {author} {\bibfnamefont {A.}~\bibnamefont
  {Peruzzo}}, \bibinfo {author} {\bibfnamefont {J.}~\bibnamefont {McClean}},
  \bibinfo {author} {\bibfnamefont {P.}~\bibnamefont {Shadbolt}}, \bibinfo
  {author} {\bibfnamefont {M.-H.}\ \bibnamefont {Yung}}, \bibinfo {author}
  {\bibfnamefont {X.-Q.}\ \bibnamefont {Zhou}}, \bibinfo {author}
  {\bibfnamefont {P.~J.}\ \bibnamefont {Love}}, \bibinfo {author}
  {\bibfnamefont {A.}~\bibnamefont {Aspuru-Guzik}},\ and\ \bibinfo {author}
  {\bibfnamefont {J.~L.}\ \bibnamefont {O'Brien}},\ }\bibfield  {title}
  {\bibinfo {title} {A variational eigenvalue solver on a photonic quantum
  processor},\ }\bibfield  {journal} {\bibinfo  {journal} {Nature
  Communications}\ }\textbf {\bibinfo {volume} {5}},\ \href
  {https://doi.org/10.1038/ncomms5213} {10.1038/ncomms5213} (\bibinfo {year}
  {2014})\BibitemShut {NoStop}%
\bibitem [{\citenamefont {Biamonte}\ \emph {et~al.}(2017)\citenamefont
  {Biamonte}, \citenamefont {Wittek}, \citenamefont {Pancotti}, \citenamefont
  {Rebentrost}, \citenamefont {Wiebe},\ and\ \citenamefont
  {Lloyd}}]{2017_biamonte_quantum_machine_learning}%
  \BibitemOpen
  \bibfield  {author} {\bibinfo {author} {\bibfnamefont {J.}~\bibnamefont
  {Biamonte}}, \bibinfo {author} {\bibfnamefont {P.}~\bibnamefont {Wittek}},
  \bibinfo {author} {\bibfnamefont {N.}~\bibnamefont {Pancotti}}, \bibinfo
  {author} {\bibfnamefont {P.}~\bibnamefont {Rebentrost}}, \bibinfo {author}
  {\bibfnamefont {N.}~\bibnamefont {Wiebe}},\ and\ \bibinfo {author}
  {\bibfnamefont {S.}~\bibnamefont {Lloyd}},\ }\bibfield  {title} {\bibinfo
  {title} {Quantum machine learning},\ }\href
  {https://doi.org/10.1038/nature23474} {\bibfield  {journal} {\bibinfo
  {journal} {Nature}\ }\textbf {\bibinfo {volume} {549}},\ \bibinfo {pages}
  {195} (\bibinfo {year} {2017})}\BibitemShut {NoStop}%
\bibitem [{\citenamefont {Cong}\ \emph {et~al.}(2019)\citenamefont {Cong},
  \citenamefont {Choi},\ and\ \citenamefont {Lukin}}]{2019_cong_quantum_cnn}%
  \BibitemOpen
  \bibfield  {author} {\bibinfo {author} {\bibfnamefont {I.}~\bibnamefont
  {Cong}}, \bibinfo {author} {\bibfnamefont {S.}~\bibnamefont {Choi}},\ and\
  \bibinfo {author} {\bibfnamefont {M.~D.}\ \bibnamefont {Lukin}},\ }\bibfield
  {title} {\bibinfo {title} {Quantum convolutional neural networks},\ }\href
  {https://doi.org/10.1038/s41567-019-0648-8} {\bibfield  {journal} {\bibinfo
  {journal} {Nature Physics}\ }\textbf {\bibinfo {volume} {15}},\ \bibinfo
  {pages} {1273} (\bibinfo {year} {2019})}\BibitemShut {NoStop}%
\bibitem [{\citenamefont {Farhi}\ \emph {et~al.}(2014)\citenamefont {Farhi},
  \citenamefont {Goldstone},\ and\ \citenamefont
  {Gutmann}}]{2014_farhi_quantum_combinatorial_opt}%
  \BibitemOpen
  \bibfield  {author} {\bibinfo {author} {\bibfnamefont {E.}~\bibnamefont
  {Farhi}}, \bibinfo {author} {\bibfnamefont {J.}~\bibnamefont {Goldstone}},\
  and\ \bibinfo {author} {\bibfnamefont {S.}~\bibnamefont {Gutmann}},\ }\href
  {https://doi.org/10.48550/ARXIV.1411.4028} {\bibinfo {title} {A quantum
  approximate optimization algorithm}} (\bibinfo {year} {2014})\BibitemShut
  {NoStop}%
\bibitem [{\citenamefont {Wang}\ \emph {et~al.}(2018)\citenamefont {Wang},
  \citenamefont {Hadfield}, \citenamefont {Jiang},\ and\ \citenamefont
  {Rieffel}}]{2018_wang_qaoa_maxcut}%
  \BibitemOpen
  \bibfield  {author} {\bibinfo {author} {\bibfnamefont {Z.}~\bibnamefont
  {Wang}}, \bibinfo {author} {\bibfnamefont {S.}~\bibnamefont {Hadfield}},
  \bibinfo {author} {\bibfnamefont {Z.}~\bibnamefont {Jiang}},\ and\ \bibinfo
  {author} {\bibfnamefont {E.~G.}\ \bibnamefont {Rieffel}},\ }\bibfield
  {title} {\bibinfo {title} {Quantum approximate optimization algorithm for
  maxcut: A fermionic view},\ }\href
  {https://doi.org/10.1103/PhysRevA.97.022304} {\bibfield  {journal} {\bibinfo
  {journal} {Phys. Rev. A}\ }\textbf {\bibinfo {volume} {97}},\ \bibinfo
  {pages} {022304} (\bibinfo {year} {2018})}\BibitemShut {NoStop}%
\bibitem [{\citenamefont {Johnson}\ \emph {et~al.}(2017)\citenamefont
  {Johnson}, \citenamefont {Romero}, \citenamefont {Olson}, \citenamefont
  {Cao},\ and\ \citenamefont
  {Aspuru-Guzik}}]{2017_johnson_var_error_correction}%
  \BibitemOpen
  \bibfield  {author} {\bibinfo {author} {\bibfnamefont {P.~D.}\ \bibnamefont
  {Johnson}}, \bibinfo {author} {\bibfnamefont {J.}~\bibnamefont {Romero}},
  \bibinfo {author} {\bibfnamefont {J.}~\bibnamefont {Olson}}, \bibinfo
  {author} {\bibfnamefont {Y.}~\bibnamefont {Cao}},\ and\ \bibinfo {author}
  {\bibfnamefont {A.}~\bibnamefont {Aspuru-Guzik}},\ }\href
  {https://doi.org/10.48550/ARXIV.1711.02249} {\bibinfo {title} {Qvector: an
  algorithm for device-tailored quantum error correction}} (\bibinfo {year}
  {2017})\BibitemShut {NoStop}%
\bibitem [{\citenamefont {Xu}\ \emph {et~al.}(2021)\citenamefont {Xu},
  \citenamefont {Benjamin},\ and\ \citenamefont
  {Yuan}}]{2021_xu_var_error_correction}%
  \BibitemOpen
  \bibfield  {author} {\bibinfo {author} {\bibfnamefont {X.}~\bibnamefont
  {Xu}}, \bibinfo {author} {\bibfnamefont {S.~C.}\ \bibnamefont {Benjamin}},\
  and\ \bibinfo {author} {\bibfnamefont {X.}~\bibnamefont {Yuan}},\ }\bibfield
  {title} {\bibinfo {title} {Variational circuit compiler for quantum error
  correction},\ }\href {https://doi.org/10.1103/PhysRevApplied.15.034068}
  {\bibfield  {journal} {\bibinfo  {journal} {Phys. Rev. Appl.}\ }\textbf
  {\bibinfo {volume} {15}},\ \bibinfo {pages} {034068} (\bibinfo {year}
  {2021})}\BibitemShut {NoStop}%
\bibitem [{\citenamefont {Khatri}\ \emph {et~al.}(2019)\citenamefont {Khatri},
  \citenamefont {LaRose}, \citenamefont {Poremba}, \citenamefont {Cincio},
  \citenamefont {Sornborger},\ and\ \citenamefont
  {Coles}}]{2019_khatri_var_quantum_compiling}%
  \BibitemOpen
  \bibfield  {author} {\bibinfo {author} {\bibfnamefont {S.}~\bibnamefont
  {Khatri}}, \bibinfo {author} {\bibfnamefont {R.}~\bibnamefont {LaRose}},
  \bibinfo {author} {\bibfnamefont {A.}~\bibnamefont {Poremba}}, \bibinfo
  {author} {\bibfnamefont {L.}~\bibnamefont {Cincio}}, \bibinfo {author}
  {\bibfnamefont {A.~T.}\ \bibnamefont {Sornborger}},\ and\ \bibinfo {author}
  {\bibfnamefont {P.~J.}\ \bibnamefont {Coles}},\ }\bibfield  {title} {\bibinfo
  {title} {Quantum-assisted quantum compiling},\ }\href
  {https://doi.org/10.22331/q-2019-05-13-140} {\bibfield  {journal} {\bibinfo
  {journal} {Quantum}\ }\textbf {\bibinfo {volume} {3}},\ \bibinfo {pages}
  {140} (\bibinfo {year} {2019})}\BibitemShut {NoStop}%
\bibitem [{\citenamefont {Sharma}\ \emph {et~al.}(2020)\citenamefont {Sharma},
  \citenamefont {Khatri}, \citenamefont {Cerezo},\ and\ \citenamefont
  {Coles}}]{2020_sharma_var_quantum_compiling}%
  \BibitemOpen
  \bibfield  {author} {\bibinfo {author} {\bibfnamefont {K.}~\bibnamefont
  {Sharma}}, \bibinfo {author} {\bibfnamefont {S.}~\bibnamefont {Khatri}},
  \bibinfo {author} {\bibfnamefont {M.}~\bibnamefont {Cerezo}},\ and\ \bibinfo
  {author} {\bibfnamefont {P.~J.}\ \bibnamefont {Coles}},\ }\bibfield  {title}
  {\bibinfo {title} {Noise resilience of variational quantum compiling},\
  }\href {https://doi.org/10.1088/1367-2630/ab784c} {\bibfield  {journal}
  {\bibinfo  {journal} {New Journal of Physics}\ }\textbf {\bibinfo {volume}
  {22}},\ \bibinfo {pages} {043006} (\bibinfo {year} {2020})}\BibitemShut
  {NoStop}%
\bibitem [{\citenamefont {Jones}\ and\ \citenamefont
  {Benjamin}(2022)}]{2022_jones_var_quantum_compilation}%
  \BibitemOpen
  \bibfield  {author} {\bibinfo {author} {\bibfnamefont {T.}~\bibnamefont
  {Jones}}\ and\ \bibinfo {author} {\bibfnamefont {S.~C.}\ \bibnamefont
  {Benjamin}},\ }\bibfield  {title} {\bibinfo {title} {Robust quantum
  compilation and circuit optimisation via energy minimisation},\ }\href
  {https://doi.org/10.22331/q-2022-01-24-628} {\bibfield  {journal} {\bibinfo
  {journal} {Quantum}\ }\textbf {\bibinfo {volume} {6}},\ \bibinfo {pages}
  {628} (\bibinfo {year} {2022})}\BibitemShut {NoStop}%
\bibitem [{\citenamefont {Li}\ and\ \citenamefont
  {Benjamin}(2017)}]{2017_ying_vqa_example}%
  \BibitemOpen
  \bibfield  {author} {\bibinfo {author} {\bibfnamefont {Y.}~\bibnamefont
  {Li}}\ and\ \bibinfo {author} {\bibfnamefont {S.~C.}\ \bibnamefont
  {Benjamin}},\ }\bibfield  {title} {\bibinfo {title} {Efficient variational
  quantum simulator incorporating active error minimization},\ }\href
  {https://doi.org/10.1103/PhysRevX.7.021050} {\bibfield  {journal} {\bibinfo
  {journal} {Phys. Rev. X}\ }\textbf {\bibinfo {volume} {7}},\ \bibinfo {pages}
  {021050} (\bibinfo {year} {2017})}\BibitemShut {NoStop}%
\bibitem [{\citenamefont {C{\^{\i}}rstoiu}\ \emph {et~al.}(2020)\citenamefont
  {C{\^{\i}}rstoiu}, \citenamefont {Holmes}, \citenamefont {Iosue},
  \citenamefont {Cincio}, \citenamefont {Coles},\ and\ \citenamefont
  {Sornborger}}]{2020_crstoiu_vff}%
  \BibitemOpen
  \bibfield  {author} {\bibinfo {author} {\bibfnamefont {C.}~\bibnamefont
  {C{\^{\i}}rstoiu}}, \bibinfo {author} {\bibfnamefont {Z.}~\bibnamefont
  {Holmes}}, \bibinfo {author} {\bibfnamefont {J.}~\bibnamefont {Iosue}},
  \bibinfo {author} {\bibfnamefont {L.}~\bibnamefont {Cincio}}, \bibinfo
  {author} {\bibfnamefont {P.~J.}\ \bibnamefont {Coles}},\ and\ \bibinfo
  {author} {\bibfnamefont {A.}~\bibnamefont {Sornborger}},\ }\bibfield  {title}
  {\bibinfo {title} {Variational fast forwarding for quantum simulation beyond
  the coherence time},\ }\bibfield  {journal} {\bibinfo  {journal} {npj Quantum
  Information}\ }\textbf {\bibinfo {volume} {6}},\ \href
  {https://doi.org/10.1038/s41534-020-00302-0} {10.1038/s41534-020-00302-0}
  (\bibinfo {year} {2020})\BibitemShut {NoStop}%
\bibitem [{\citenamefont {Yao}\ \emph {et~al.}(2021)\citenamefont {Yao},
  \citenamefont {Gomes}, \citenamefont {Zhang}, \citenamefont {Wang},
  \citenamefont {Ho}, \citenamefont {Iadecola},\ and\ \citenamefont
  {Orth}}]{2021_yao_adaptive_tdva}%
  \BibitemOpen
  \bibfield  {author} {\bibinfo {author} {\bibfnamefont {Y.-X.}\ \bibnamefont
  {Yao}}, \bibinfo {author} {\bibfnamefont {N.}~\bibnamefont {Gomes}}, \bibinfo
  {author} {\bibfnamefont {F.}~\bibnamefont {Zhang}}, \bibinfo {author}
  {\bibfnamefont {C.-Z.}\ \bibnamefont {Wang}}, \bibinfo {author}
  {\bibfnamefont {K.-M.}\ \bibnamefont {Ho}}, \bibinfo {author} {\bibfnamefont
  {T.}~\bibnamefont {Iadecola}},\ and\ \bibinfo {author} {\bibfnamefont
  {P.~P.}\ \bibnamefont {Orth}},\ }\bibfield  {title} {\bibinfo {title}
  {Adaptive variational quantum dynamics simulations},\ }\href
  {https://doi.org/10.1103/PRXQuantum.2.030307} {\bibfield  {journal} {\bibinfo
   {journal} {PRX Quantum}\ }\textbf {\bibinfo {volume} {2}},\ \bibinfo {pages}
  {030307} (\bibinfo {year} {2021})}\BibitemShut {NoStop}%
\bibitem [{\citenamefont {Lin}\ \emph {et~al.}(2021)\citenamefont {Lin},
  \citenamefont {Dilip}, \citenamefont {Green}, \citenamefont {Smith},\ and\
  \citenamefont {Pollmann}}]{Hsuan_2021}%
  \BibitemOpen
  \bibfield  {author} {\bibinfo {author} {\bibfnamefont {S.-H.}\ \bibnamefont
  {Lin}}, \bibinfo {author} {\bibfnamefont {R.}~\bibnamefont {Dilip}}, \bibinfo
  {author} {\bibfnamefont {A.~G.}\ \bibnamefont {Green}}, \bibinfo {author}
  {\bibfnamefont {A.}~\bibnamefont {Smith}},\ and\ \bibinfo {author}
  {\bibfnamefont {F.}~\bibnamefont {Pollmann}},\ }\bibfield  {title} {\bibinfo
  {title} {Real- and imaginary-time evolution with compressed quantum
  circuits},\ }\href {https://doi.org/10.1103/PRXQuantum.2.010342} {\bibfield
  {journal} {\bibinfo  {journal} {PRX Quantum}\ }\textbf {\bibinfo {volume}
  {2}},\ \bibinfo {pages} {010342} (\bibinfo {year} {2021})}\BibitemShut
  {NoStop}%
\bibitem [{\citenamefont {Barratt}\ \emph {et~al.}(2021)\citenamefont
  {Barratt}, \citenamefont {Dborin}, \citenamefont {Bal}, \citenamefont
  {Stojevic}, \citenamefont {Pollmann},\ and\ \citenamefont
  {Green}}]{Barratt_2021}%
  \BibitemOpen
  \bibfield  {author} {\bibinfo {author} {\bibfnamefont {F.}~\bibnamefont
  {Barratt}}, \bibinfo {author} {\bibfnamefont {J.}~\bibnamefont {Dborin}},
  \bibinfo {author} {\bibfnamefont {M.}~\bibnamefont {Bal}}, \bibinfo {author}
  {\bibfnamefont {V.}~\bibnamefont {Stojevic}}, \bibinfo {author}
  {\bibfnamefont {F.}~\bibnamefont {Pollmann}},\ and\ \bibinfo {author}
  {\bibfnamefont {A.~G.}\ \bibnamefont {Green}},\ }\bibfield  {title} {\bibinfo
  {title} {{Parallel quantum simulation of large systems on small NISQ
  computers}},\ }\href {https://doi.org/10.1038/s41534-021-00420-3} {\bibfield
  {journal} {\bibinfo  {journal} {npj Quantum Information}\ }\textbf {\bibinfo
  {volume} {7}},\ \bibinfo {pages} {79} (\bibinfo {year} {2021})}\BibitemShut
  {NoStop}%
\bibitem [{\citenamefont {Barison}\ \emph {et~al.}(2021)\citenamefont
  {Barison}, \citenamefont {Vicentini},\ and\ \citenamefont
  {Carleo}}]{2021_barison_p-vqd}%
  \BibitemOpen
  \bibfield  {author} {\bibinfo {author} {\bibfnamefont {S.}~\bibnamefont
  {Barison}}, \bibinfo {author} {\bibfnamefont {F.}~\bibnamefont {Vicentini}},\
  and\ \bibinfo {author} {\bibfnamefont {G.}~\bibnamefont {Carleo}},\
  }\bibfield  {title} {\bibinfo {title} {An efficient quantum algorithm for the
  time evolution of parameterized circuits},\ }\href
  {https://doi.org/10.22331/q-2021-07-28-512} {\bibfield  {journal} {\bibinfo
  {journal} {Quantum}\ }\textbf {\bibinfo {volume} {5}},\ \bibinfo {pages}
  {512} (\bibinfo {year} {2021})}\BibitemShut {NoStop}%
\bibitem [{\citenamefont {Berthusen}\ \emph {et~al.}(2022)\citenamefont
  {Berthusen}, \citenamefont {Trevisan}, \citenamefont {Iadecola},\ and\
  \citenamefont {Orth}}]{2022_berthusen_dynamics_on_hardware}%
  \BibitemOpen
  \bibfield  {author} {\bibinfo {author} {\bibfnamefont {N.~F.}\ \bibnamefont
  {Berthusen}}, \bibinfo {author} {\bibfnamefont {T.~V.}\ \bibnamefont
  {Trevisan}}, \bibinfo {author} {\bibfnamefont {T.}~\bibnamefont {Iadecola}},\
  and\ \bibinfo {author} {\bibfnamefont {P.~P.}\ \bibnamefont {Orth}},\
  }\bibfield  {title} {\bibinfo {title} {Quantum dynamics simulations beyond
  the coherence time on noisy intermediate-scale quantum hardware by
  variational trotter compression},\ }\bibfield  {journal} {\bibinfo  {journal}
  {Physical Review Research}\ }\textbf {\bibinfo {volume} {4}},\ \href
  {https://doi.org/10.1103/physrevresearch.4.023097}
  {10.1103/physrevresearch.4.023097} (\bibinfo {year} {2022})\BibitemShut
  {NoStop}%
\bibitem [{\citenamefont {Barison}\ \emph {et~al.}(2022)\citenamefont
  {Barison}, \citenamefont {Vicentini}, \citenamefont {Cirac},\ and\
  \citenamefont {Carleo}}]{2022_barison_vfk}%
  \BibitemOpen
  \bibfield  {author} {\bibinfo {author} {\bibfnamefont {S.}~\bibnamefont
  {Barison}}, \bibinfo {author} {\bibfnamefont {F.}~\bibnamefont {Vicentini}},
  \bibinfo {author} {\bibfnamefont {I.}~\bibnamefont {Cirac}},\ and\ \bibinfo
  {author} {\bibfnamefont {G.}~\bibnamefont {Carleo}},\ }\bibfield  {title}
  {\bibinfo {title} {Variational dynamics as a ground-state problem on a
  quantum computer},\ }\href {https://doi.org/10.1103/PhysRevResearch.4.043161}
  {\bibfield  {journal} {\bibinfo  {journal} {Phys. Rev. Res.}\ }\textbf
  {\bibinfo {volume} {4}},\ \bibinfo {pages} {043161} (\bibinfo {year}
  {2022})}\BibitemShut {NoStop}%
\bibitem [{\citenamefont {Miessen}\ \emph {et~al.}(2023)\citenamefont
  {Miessen}, \citenamefont {Ollitrault}, \citenamefont {Tacchino},\ and\
  \citenamefont {Tavernelli}}]{Miessen2023_te}%
  \BibitemOpen
  \bibfield  {author} {\bibinfo {author} {\bibfnamefont {A.}~\bibnamefont
  {Miessen}}, \bibinfo {author} {\bibfnamefont {P.~J.}\ \bibnamefont
  {Ollitrault}}, \bibinfo {author} {\bibfnamefont {F.}~\bibnamefont
  {Tacchino}},\ and\ \bibinfo {author} {\bibfnamefont {I.}~\bibnamefont
  {Tavernelli}},\ }\bibfield  {title} {\bibinfo {title} {{Quantum algorithms
  for quantum dynamics}},\ }\href {https://doi.org/10.1038/s43588-022-00374-2}
  {\bibfield  {journal} {\bibinfo  {journal} {Nature Computational Science}\
  }\textbf {\bibinfo {volume} {3}},\ \bibinfo {pages} {25} (\bibinfo {year}
  {2023})}\BibitemShut {NoStop}%
\bibitem [{\citenamefont {Trotter}(1959)}]{1959_trotter}%
  \BibitemOpen
  \bibfield  {author} {\bibinfo {author} {\bibfnamefont {H.~F.}\ \bibnamefont
  {Trotter}},\ }\bibfield  {title} {\bibinfo {title} {On the product of
  semi-groups of operators},\ }\href {http://www.jstor.org/stable/2033649}
  {\bibfield  {journal} {\bibinfo  {journal} {Proceedings of the American
  Mathematical Society}\ }\textbf {\bibinfo {volume} {10}},\ \bibinfo {pages}
  {545} (\bibinfo {year} {1959})}\BibitemShut {NoStop}%
\bibitem [{\citenamefont {Suzuki}(1976)}]{1976_suzuki}%
  \BibitemOpen
  \bibfield  {author} {\bibinfo {author} {\bibfnamefont {M.}~\bibnamefont
  {Suzuki}},\ }\bibfield  {title} {\bibinfo {title} {Generalized trotter's
  formula and systematic approximants of exponential operators and inner
  derivations with applications to many-body problems},\ }\href@noop {}
  {\bibfield  {journal} {\bibinfo  {journal} {Communications in Mathematical
  Physics}\ }\textbf {\bibinfo {volume} {51}},\ \bibinfo {pages} {183}
  (\bibinfo {year} {1976})}\BibitemShut {NoStop}%
\bibitem [{\citenamefont {Abrams}\ and\ \citenamefont
  {Lloyd}(1997)}]{1997_abrams_trotter_for_simulation}%
  \BibitemOpen
  \bibfield  {author} {\bibinfo {author} {\bibfnamefont {D.~S.}\ \bibnamefont
  {Abrams}}\ and\ \bibinfo {author} {\bibfnamefont {S.}~\bibnamefont {Lloyd}},\
  }\bibfield  {title} {\bibinfo {title} {Simulation of many-body fermi systems
  on a universal quantum computer},\ }\href
  {https://doi.org/10.1103/PhysRevLett.79.2586} {\bibfield  {journal} {\bibinfo
   {journal} {Phys. Rev. Lett.}\ }\textbf {\bibinfo {volume} {79}},\ \bibinfo
  {pages} {2586} (\bibinfo {year} {1997})}\BibitemShut {NoStop}%
\bibitem [{\citenamefont {Ortiz}\ \emph {et~al.}(2001)\citenamefont {Ortiz},
  \citenamefont {Gubernatis}, \citenamefont {Knill},\ and\ \citenamefont
  {Laflamme}}]{2001_ortiz_trotter_for_simulation}%
  \BibitemOpen
  \bibfield  {author} {\bibinfo {author} {\bibfnamefont {G.}~\bibnamefont
  {Ortiz}}, \bibinfo {author} {\bibfnamefont {J.~E.}\ \bibnamefont
  {Gubernatis}}, \bibinfo {author} {\bibfnamefont {E.}~\bibnamefont {Knill}},\
  and\ \bibinfo {author} {\bibfnamefont {R.}~\bibnamefont {Laflamme}},\
  }\bibfield  {title} {\bibinfo {title} {Quantum algorithms for fermionic
  simulations},\ }\href {https://doi.org/10.1103/PhysRevA.64.022319} {\bibfield
   {journal} {\bibinfo  {journal} {Phys. Rev. A}\ }\textbf {\bibinfo {volume}
  {64}},\ \bibinfo {pages} {022319} (\bibinfo {year} {2001})}\BibitemShut
  {NoStop}%
\bibitem [{\citenamefont {Tacchino}\ \emph {et~al.}(2019)\citenamefont
  {Tacchino}, \citenamefont {Chiesa}, \citenamefont {Carretta},\ and\
  \citenamefont {Gerace}}]{Tacchino_2019}%
  \BibitemOpen
  \bibfield  {author} {\bibinfo {author} {\bibfnamefont {F.}~\bibnamefont
  {Tacchino}}, \bibinfo {author} {\bibfnamefont {A.}~\bibnamefont {Chiesa}},
  \bibinfo {author} {\bibfnamefont {S.}~\bibnamefont {Carretta}},\ and\
  \bibinfo {author} {\bibfnamefont {D.}~\bibnamefont {Gerace}},\ }\bibfield
  {title} {\bibinfo {title} {Quantum computers as universal quantum simulators:
  State-of-the-art and perspectives},\ }\href
  {https://doi.org/10.1002/qute.201900052} {\bibfield  {journal} {\bibinfo
  {journal} {Advanced Quantum Technologies}\ }\textbf {\bibinfo {volume} {3}},\
  \bibinfo {pages} {1900052} (\bibinfo {year} {2019})}\BibitemShut {NoStop}%
\bibitem [{\citenamefont {Grimsley}\ \emph {et~al.}(2019)\citenamefont
  {Grimsley}, \citenamefont {Economou}, \citenamefont {Barnes},\ and\
  \citenamefont {Mayhall}}]{2019_grimsley_adapt-vqe}%
  \BibitemOpen
  \bibfield  {author} {\bibinfo {author} {\bibfnamefont {H.~R.}\ \bibnamefont
  {Grimsley}}, \bibinfo {author} {\bibfnamefont {S.~E.}\ \bibnamefont
  {Economou}}, \bibinfo {author} {\bibfnamefont {E.}~\bibnamefont {Barnes}},\
  and\ \bibinfo {author} {\bibfnamefont {N.~J.}\ \bibnamefont {Mayhall}},\
  }\bibfield  {title} {\bibinfo {title} {An adaptive variational algorithm for
  exact molecular simulations on a quantum computer},\ }\href
  {https://doi.org/10.1038/s41467-019-10988-2} {\bibfield  {journal} {\bibinfo
  {journal} {Nature Communications}\ }\textbf {\bibinfo {volume} {10}},\
  \bibinfo {pages} {3007} (\bibinfo {year} {2019})}\BibitemShut {NoStop}%
\bibitem [{\citenamefont {Tang}\ \emph {et~al.}(2021)\citenamefont {Tang},
  \citenamefont {Shkolnikov}, \citenamefont {Barron}, \citenamefont {Grimsley},
  \citenamefont {Mayhall}, \citenamefont {Barnes},\ and\ \citenamefont
  {Economou}}]{2021_tang_qubit_adapt_vqe}%
  \BibitemOpen
  \bibfield  {author} {\bibinfo {author} {\bibfnamefont {H.~L.}\ \bibnamefont
  {Tang}}, \bibinfo {author} {\bibfnamefont {V.}~\bibnamefont {Shkolnikov}},
  \bibinfo {author} {\bibfnamefont {G.~S.}\ \bibnamefont {Barron}}, \bibinfo
  {author} {\bibfnamefont {H.~R.}\ \bibnamefont {Grimsley}}, \bibinfo {author}
  {\bibfnamefont {N.~J.}\ \bibnamefont {Mayhall}}, \bibinfo {author}
  {\bibfnamefont {E.}~\bibnamefont {Barnes}},\ and\ \bibinfo {author}
  {\bibfnamefont {S.~E.}\ \bibnamefont {Economou}},\ }\bibfield  {title}
  {\bibinfo {title} {Qubit-{ADAPT}-{VQE}: An adaptive algorithm for
  constructing hardware-efficient ansätze on a quantum processor},\ }\bibfield
   {journal} {\bibinfo  {journal} {{PRX} Quantum}\ }\textbf {\bibinfo {volume}
  {2}},\ \href {https://doi.org/10.1103/prxquantum.2.020310}
  {10.1103/prxquantum.2.020310} (\bibinfo {year} {2021})\BibitemShut {NoStop}%
\bibitem [{\citenamefont {Van~Dyke}\ \emph {et~al.}(2022)\citenamefont
  {Van~Dyke}, \citenamefont {Barron}, \citenamefont {Mayhall}, \citenamefont
  {Barnes},\ and\ \citenamefont {Economou}}]{2022_van_dyke_pool_tiling}%
  \BibitemOpen
  \bibfield  {author} {\bibinfo {author} {\bibfnamefont {J.~S.}\ \bibnamefont
  {Van~Dyke}}, \bibinfo {author} {\bibfnamefont {G.~S.}\ \bibnamefont
  {Barron}}, \bibinfo {author} {\bibfnamefont {N.~J.}\ \bibnamefont {Mayhall}},
  \bibinfo {author} {\bibfnamefont {E.}~\bibnamefont {Barnes}},\ and\ \bibinfo
  {author} {\bibfnamefont {S.~E.}\ \bibnamefont {Economou}},\ }\href
  {https://doi.org/10.48550/ARXIV.2206.14215} {\bibinfo {title} {Scaling
  adaptive quantum simulation algorithms via operator pool tiling}} (\bibinfo
  {year} {2022})\BibitemShut {NoStop}%
\bibitem [{\citenamefont {Anastasiou}\ \emph {et~al.}(2022)\citenamefont
  {Anastasiou}, \citenamefont {Chen}, \citenamefont {Mayhall}, \citenamefont
  {Barnes},\ and\ \citenamefont {Economou}}]{2022_economou_tetris}%
  \BibitemOpen
  \bibfield  {author} {\bibinfo {author} {\bibfnamefont {P.~G.}\ \bibnamefont
  {Anastasiou}}, \bibinfo {author} {\bibfnamefont {Y.}~\bibnamefont {Chen}},
  \bibinfo {author} {\bibfnamefont {N.~J.}\ \bibnamefont {Mayhall}}, \bibinfo
  {author} {\bibfnamefont {E.}~\bibnamefont {Barnes}},\ and\ \bibinfo {author}
  {\bibfnamefont {S.~E.}\ \bibnamefont {Economou}},\ }\href
  {https://doi.org/10.48550/ARXIV.2209.10562} {\bibinfo {title}
  {Tetris-adapt-vqe: An adaptive algorithm that yields shallower, denser
  circuit ansätze}} (\bibinfo {year} {2022})\BibitemShut {NoStop}%
\bibitem [{\citenamefont {Gomes}\ \emph {et~al.}(2023)\citenamefont {Gomes},
  \citenamefont {Williams-Young},\ and\ \citenamefont
  {de~Jong}}]{Niladri_2023}%
  \BibitemOpen
  \bibfield  {author} {\bibinfo {author} {\bibfnamefont {N.}~\bibnamefont
  {Gomes}}, \bibinfo {author} {\bibfnamefont {D.~B.}\ \bibnamefont
  {Williams-Young}},\ and\ \bibinfo {author} {\bibfnamefont {W.~A.}\
  \bibnamefont {de~Jong}},\ }\href {https://doi.org/10.48550/ARXIV.2302.03093}
  {\bibinfo {title} {Computing the many-body green's function with adaptive
  variational quantum dynamics}} (\bibinfo {year} {2023})\BibitemShut {NoStop}%
\bibitem [{\citenamefont {Anastasiou}\ \emph {et~al.}(2023)\citenamefont
  {Anastasiou}, \citenamefont {Mayhall}, \citenamefont {Barnes},\ and\
  \citenamefont {Economou}}]{anastasiou2023really}%
  \BibitemOpen
  \bibfield  {author} {\bibinfo {author} {\bibfnamefont {P.~G.}\ \bibnamefont
  {Anastasiou}}, \bibinfo {author} {\bibfnamefont {N.~J.}\ \bibnamefont
  {Mayhall}}, \bibinfo {author} {\bibfnamefont {E.}~\bibnamefont {Barnes}},\
  and\ \bibinfo {author} {\bibfnamefont {S.~E.}\ \bibnamefont {Economou}},\
  }\href@noop {} {\bibinfo {title} {How to really measure operator gradients in
  adapt-vqe}} (\bibinfo {year} {2023}),\ \Eprint
  {https://arxiv.org/abs/2306.03227} {arXiv:2306.03227 [quant-ph]} \BibitemShut
  {NoStop}%
\bibitem [{\citenamefont {Mari}\ \emph {et~al.}(2021)\citenamefont {Mari},
  \citenamefont {Bromley},\ and\ \citenamefont
  {Killoran}}]{2021_mari_param_shift_rule}%
  \BibitemOpen
  \bibfield  {author} {\bibinfo {author} {\bibfnamefont {A.}~\bibnamefont
  {Mari}}, \bibinfo {author} {\bibfnamefont {T.~R.}\ \bibnamefont {Bromley}},\
  and\ \bibinfo {author} {\bibfnamefont {N.}~\bibnamefont {Killoran}},\
  }\bibfield  {title} {\bibinfo {title} {Estimating the gradient and
  higher-order derivatives on quantum hardware},\ }\bibfield  {journal}
  {\bibinfo  {journal} {Physical Review A}\ }\textbf {\bibinfo {volume}
  {103}},\ \href {https://doi.org/10.1103/physreva.103.012405}
  {10.1103/physreva.103.012405} (\bibinfo {year} {2021})\BibitemShut {NoStop}%
\bibitem [{\citenamefont {Shkolnikov}\ \emph {et~al.}(2021)\citenamefont
  {Shkolnikov}, \citenamefont {Mayhall}, \citenamefont {Economou},\ and\
  \citenamefont {Barnes}}]{shkolnikov2021avoiding}%
  \BibitemOpen
  \bibfield  {author} {\bibinfo {author} {\bibfnamefont {V.~O.}\ \bibnamefont
  {Shkolnikov}}, \bibinfo {author} {\bibfnamefont {N.~J.}\ \bibnamefont
  {Mayhall}}, \bibinfo {author} {\bibfnamefont {S.~E.}\ \bibnamefont
  {Economou}},\ and\ \bibinfo {author} {\bibfnamefont {E.}~\bibnamefont
  {Barnes}},\ }\href@noop {} {\bibinfo {title} {Avoiding symmetry roadblocks
  and minimizing the measurement overhead of adaptive variational quantum
  eigensolvers}} (\bibinfo {year} {2021}),\ \Eprint
  {https://arxiv.org/abs/2109.05340} {arXiv:2109.05340 [quant-ph]} \BibitemShut
  {NoStop}%
\bibitem [{\citenamefont {Yordanov}\ \emph {et~al.}(2021)\citenamefont
  {Yordanov}, \citenamefont {Armaos}, \citenamefont {Barnes},\ and\
  \citenamefont {Arvidsson-Shukur}}]{2021_yordanov_adapt}%
  \BibitemOpen
  \bibfield  {author} {\bibinfo {author} {\bibfnamefont {Y.~S.}\ \bibnamefont
  {Yordanov}}, \bibinfo {author} {\bibfnamefont {V.}~\bibnamefont {Armaos}},
  \bibinfo {author} {\bibfnamefont {C.~H.~W.}\ \bibnamefont {Barnes}},\ and\
  \bibinfo {author} {\bibfnamefont {D.~R.~M.}\ \bibnamefont
  {Arvidsson-Shukur}},\ }\bibfield  {title} {\bibinfo {title}
  {{Qubit-excitation-based adaptive variational quantum eigensolver}},\ }\href
  {https://doi.org/10.1038/s42005-021-00730-0} {\bibfield  {journal} {\bibinfo
  {journal} {Communications Physics}\ }\textbf {\bibinfo {volume} {4}},\
  \bibinfo {pages} {228} (\bibinfo {year} {2021})}\BibitemShut {NoStop}%
\bibitem [{\citenamefont {Temme}\ \emph {et~al.}(2017)\citenamefont {Temme},
  \citenamefont {Bravyi},\ and\ \citenamefont {Gambetta}}]{Temme_2017_zne}%
  \BibitemOpen
  \bibfield  {author} {\bibinfo {author} {\bibfnamefont {K.}~\bibnamefont
  {Temme}}, \bibinfo {author} {\bibfnamefont {S.}~\bibnamefont {Bravyi}},\ and\
  \bibinfo {author} {\bibfnamefont {J.~M.}\ \bibnamefont {Gambetta}},\
  }\bibfield  {title} {\bibinfo {title} {Error mitigation for short-depth
  quantum circuits},\ }\href {https://doi.org/10.1103/PhysRevLett.119.180509}
  {\bibfield  {journal} {\bibinfo  {journal} {Phys. Rev. Lett.}\ }\textbf
  {\bibinfo {volume} {119}},\ \bibinfo {pages} {180509} (\bibinfo {year}
  {2017})}\BibitemShut {NoStop}%
\bibitem [{\citenamefont {tA~v}\ \emph {et~al.}(2021)\citenamefont {tA~v} \emph
  {et~al.}}]{qiskit}%
  \BibitemOpen
  \bibfield  {author} {\bibinfo {author} {\bibfnamefont {A.}~\bibnamefont
  {tA~v}} \emph {et~al.},\ }\href {https://doi.org/10.5281/zenodo.2573505}
  {\bibinfo {title} {Qiskit: An open-source framework for quantum computing}}
  (\bibinfo {year} {2021})\BibitemShut {NoStop}%
\bibitem [{\citenamefont {van~den Berg}\ \emph {et~al.}(2023)\citenamefont
  {van~den Berg}, \citenamefont {Minev}, \citenamefont {Kandala},\ and\
  \citenamefont {Temme}}]{VandenBerg2023}%
  \BibitemOpen
  \bibfield  {author} {\bibinfo {author} {\bibfnamefont {E.}~\bibnamefont
  {van~den Berg}}, \bibinfo {author} {\bibfnamefont {Z.~K.}\ \bibnamefont
  {Minev}}, \bibinfo {author} {\bibfnamefont {A.}~\bibnamefont {Kandala}},\
  and\ \bibinfo {author} {\bibfnamefont {K.}~\bibnamefont {Temme}},\ }\bibfield
   {title} {\bibinfo {title} {{Probabilistic error cancellation with sparse
  Pauli–Lindblad models on noisy quantum processors}},\ }\bibfield  {journal}
  {\bibinfo  {journal} {Nature Physics}\ }\href
  {https://doi.org/10.1038/s41567-023-02042-2} {10.1038/s41567-023-02042-2}
  (\bibinfo {year} {2023})\BibitemShut {NoStop}%
\bibitem [{\citenamefont {E.P.~Jornan}(1993)}]{Jordan93}%
  \BibitemOpen
  \bibfield  {author} {\bibinfo {author} {\bibfnamefont {E.~W.}\ \bibnamefont
  {E.P.~Jornan}},\ }\href
  {https://doi.org/https://doi.org/10.1007/978-3-662-02781-3} {\emph {\bibinfo
  {title} {The Collected Works of {Eugene Paul Wigner}}}}\ (\bibinfo
  {publisher} {Springer, Berlin, Heidelberg},\ \bibinfo {year}
  {1993})\BibitemShut {NoStop}%
\bibitem [{\citenamefont {Bravyi}\ and\ \citenamefont
  {Kitaev}(2002)}]{Bravyi02_bk}%
  \BibitemOpen
  \bibfield  {author} {\bibinfo {author} {\bibfnamefont {S.~B.}\ \bibnamefont
  {Bravyi}}\ and\ \bibinfo {author} {\bibfnamefont {A.~Y.}\ \bibnamefont
  {Kitaev}},\ }\bibfield  {title} {\bibinfo {title} {Fermionic quantum
  computation},\ }\href {https://doi.org/10.1006/aphy.2002.6254} {\bibfield
  {journal} {\bibinfo  {journal} {Annals of Physics}\ }\textbf {\bibinfo
  {volume} {298}},\ \bibinfo {pages} {210} (\bibinfo {year}
  {2002})}\BibitemShut {NoStop}%
\bibitem [{\citenamefont {Verstraete}\ and\ \citenamefont
  {Cirac}(2005)}]{Verstraete_2005}%
  \BibitemOpen
  \bibfield  {author} {\bibinfo {author} {\bibfnamefont {F.}~\bibnamefont
  {Verstraete}}\ and\ \bibinfo {author} {\bibfnamefont {J.~I.}\ \bibnamefont
  {Cirac}},\ }\bibfield  {title} {\bibinfo {title} {Mapping local hamiltonians
  of fermions to local hamiltonians of spins},\ }\href
  {https://doi.org/10.1088/1742-5468/2005/09/P09012} {\bibfield  {journal}
  {\bibinfo  {journal} {Journal of Statistical Mechanics: Theory and
  Experiment}\ }\textbf {\bibinfo {volume} {2005}},\ \bibinfo {pages} {P09012}
  (\bibinfo {year} {2005})}\BibitemShut {NoStop}%
\bibitem [{\citenamefont {Whitfield}\ \emph {et~al.}(2016)\citenamefont
  {Whitfield}, \citenamefont {Havl\'{\i}\ifmmode~\check{c}\else \v{c}\fi{}ek},\
  and\ \citenamefont {Troyer}}]{Whitfield_2016}%
  \BibitemOpen
  \bibfield  {author} {\bibinfo {author} {\bibfnamefont {J.~D.}\ \bibnamefont
  {Whitfield}}, \bibinfo {author} {\bibfnamefont {V.~c.~v.}\ \bibnamefont
  {Havl\'{\i}\ifmmode~\check{c}\else \v{c}\fi{}ek}},\ and\ \bibinfo {author}
  {\bibfnamefont {M.}~\bibnamefont {Troyer}},\ }\bibfield  {title} {\bibinfo
  {title} {Local spin operators for fermion simulations},\ }\href
  {https://doi.org/10.1103/PhysRevA.94.030301} {\bibfield  {journal} {\bibinfo
  {journal} {Phys. Rev. A}\ }\textbf {\bibinfo {volume} {94}},\ \bibinfo
  {pages} {030301} (\bibinfo {year} {2016})}\BibitemShut {NoStop}%
\bibitem [{\citenamefont {Setia}\ \emph {et~al.}(2019)\citenamefont {Setia},
  \citenamefont {Bravyi}, \citenamefont {Mezzacapo},\ and\ \citenamefont
  {Whitfield}}]{Setia_2019}%
  \BibitemOpen
  \bibfield  {author} {\bibinfo {author} {\bibfnamefont {K.}~\bibnamefont
  {Setia}}, \bibinfo {author} {\bibfnamefont {S.}~\bibnamefont {Bravyi}},
  \bibinfo {author} {\bibfnamefont {A.}~\bibnamefont {Mezzacapo}},\ and\
  \bibinfo {author} {\bibfnamefont {J.~D.}\ \bibnamefont {Whitfield}},\
  }\bibfield  {title} {\bibinfo {title} {Superfast encodings for fermionic
  quantum simulation},\ }\href
  {https://doi.org/10.1103/PhysRevResearch.1.033033} {\bibfield  {journal}
  {\bibinfo  {journal} {Phys. Rev. Res.}\ }\textbf {\bibinfo {volume} {1}},\
  \bibinfo {pages} {033033} (\bibinfo {year} {2019})}\BibitemShut {NoStop}%
\bibitem [{\citenamefont {Chen}\ and\ \citenamefont {Xu}(2022)}]{Chen_2022}%
  \BibitemOpen
  \bibfield  {author} {\bibinfo {author} {\bibfnamefont {Y.-A.}\ \bibnamefont
  {Chen}}\ and\ \bibinfo {author} {\bibfnamefont {Y.}~\bibnamefont {Xu}},\
  }\href {https://doi.org/10.48550/ARXIV.2201.05153} {\bibinfo {title}
  {Equivalence between fermion-to-qubit mappings in two spatial dimensions}}
  (\bibinfo {year} {2022})\BibitemShut {NoStop}%
\bibitem [{\citenamefont {Nys}\ and\ \citenamefont {Carleo}(2023)}]{Nys_2023}%
  \BibitemOpen
  \bibfield  {author} {\bibinfo {author} {\bibfnamefont {J.}~\bibnamefont
  {Nys}}\ and\ \bibinfo {author} {\bibfnamefont {G.}~\bibnamefont {Carleo}},\
  }\bibfield  {title} {\bibinfo {title} {Quantum circuits for solving local
  fermion-to-qubit mappings},\ }\href
  {https://doi.org/10.22331/q-2023-02-21-930} {\bibfield  {journal} {\bibinfo
  {journal} {{Quantum}}\ }\textbf {\bibinfo {volume} {7}},\ \bibinfo {pages}
  {930} (\bibinfo {year} {2023})}\BibitemShut {NoStop}%
\bibitem [{\citenamefont {Linteau}(2022)}]{github}%
  \BibitemOpen
  \bibfield  {author} {\bibinfo {author} {\bibfnamefont {D.}~\bibnamefont
  {Linteau}},\ }\href@noop {} {\bibinfo {title} {adaptive-pvqd}},\ \bibinfo
  {howpublished} {\url{https://github.com/dalin27/adaptive-pvqd}} (\bibinfo
  {year} {2022})\BibitemShut {NoStop}%
\bibitem [{\citenamefont {Johansson}\ \emph {et~al.}(2012)\citenamefont
  {Johansson}, \citenamefont {Nation},\ and\ \citenamefont {Nori}}]{qutip}%
  \BibitemOpen
  \bibfield  {author} {\bibinfo {author} {\bibfnamefont {J.~R.}\ \bibnamefont
  {Johansson}}, \bibinfo {author} {\bibfnamefont {P.~D.}\ \bibnamefont
  {Nation}},\ and\ \bibinfo {author} {\bibfnamefont {F.}~\bibnamefont {Nori}},\
  }\bibfield  {title} {\bibinfo {title} {Qutip: An open-source python framework
  for the dynamics of open quantum systems},\ }\bibfield  {journal} {\bibinfo
  {journal} {Comp. Phys. Comm.}\ }\textbf {\bibinfo {volume} {183}},\ \href
  {https://doi.org/10.1016/j.cpc.2012.02.021} {10.1016/j.cpc.2012.02.021}
  (\bibinfo {year} {2012})\BibitemShut {NoStop}%
\bibitem [{\citenamefont {Kingma}\ and\ \citenamefont
  {Ba}(2014)}]{2014_kingma_adam_optimizer}%
  \BibitemOpen
  \bibfield  {author} {\bibinfo {author} {\bibfnamefont {D.~P.}\ \bibnamefont
  {Kingma}}\ and\ \bibinfo {author} {\bibfnamefont {J.}~\bibnamefont {Ba}},\
  }\href {https://doi.org/10.48550/ARXIV.1412.6980} {\bibinfo {title} {Adam: A
  method for stochastic optimization}} (\bibinfo {year} {2014})\BibitemShut
  {NoStop}%
\end{thebibliography}
\end{document}